\documentclass[useAMS,usenatbib]{mn2e}
\usepackage{graphicx,xspace,longtable}
\usepackage[usenames]{color}
\usepackage[toc,page]{appendix}
\usepackage{rotate,amssymb,amsmath,shadow,bezier,curves}

\usepackage{appendix}

\newcommand{\beq}{\begin{equation}}
\newcommand{\eeq}{\end{equation}}
\newcommand{\pa}{\partial}

\title[Distortion of Infall Regions in Redshift Space]{Distortion of Infall Regions in Redshift Space-I}
\author[Abdullah M. H., Praton E. A.\& Ali G. B.]{Mohamed H. Abdullah$^1$, Elizabeth A. Praton$^2$ \& Gamal B. Ali$^1$\\
$^1$ Department of Astronomy, National Research Institute of Astronomy and Geophysics, Egypt\\
$^2$ Department of Physics and Astronomy, Franklin \& Marshall College, USA}

\begin{document}
\maketitle

\begin{abstract}

We show that spherical infall models (SIMs) can better describe some galaxy clusters in redshift slice space than in traditional axially-convolved projection space. This is because in SIM, the presence of transverse motion between cluster and observer, and/or shear flow about the cluster (such as rotation), causes the infall artifact to tilt, obscuring the characteristic two-trumpet profile; and some clusters resemble such tilted artifacts.  

We illustrate the disadvantages of applying SIM to convolved data and, as an alternative, introduce a method fitting a tilted 2D envelope to determine a 3D envelope. We also introduce a fitting algorithm and test it on toy SIM simulations as well as three clusters (Virgo, A1459, and A1066). We derive relations useful for using the tilt and width-to-length ratio of the fitted envelopes to analyze peculiar velocities. We apply them to our three clusters as a demonstration. We find that transverse motion between cluster and observer can be ruled out as sole cause of the observed tilts, and that a multi-cluster study could be a feasible way to find our infall toward Virgo cluster.

\end{abstract}

\section{Introduction}
\label{sec:Intro}

Studying the properties of galaxy clusters plays an important role in investigating formation of large scale structure and constraining cosmological parameters. The peculiar velocities associated with any inhomogeneous structure introduce a distortion in the redshift mapping, and any analysis must take this into account. In this paper, we are interested in the distortion of the infall region surrounding a cluster core, so hereafter we will use the term galaxy cluster to mean everything inside the turnaround (which separates galaxies moving inward from those moving outward). 

Galaxy clusters have been explored in redshift space by a number of authors, e.g., Kent \& Gunn (1982), Reg\H{o}s \& Geller (1989), hereafter RG89, Geller, Diaferio \& Kurtz (1999), Drinkwater, Gregg \& Colless (2001), Reisenegger et al. (2000), and Abdullah et al. (2011), hereafter AA11. One common approach has been to use a spherical infall model, hereafter SIM, to try to fit an envelope in redshift space. 

There are many different types of SIMs (e.g., Gunn \& Gott 1972; Schechter 1980; RG89; Praton \& Schneider 1994, hereafter PS94) but in all of them, the infall region distorts in redshift space into a structure which has the form of two trumpet horns glued face to face (Kaiser 1987). This artifact arises because the positions of galaxies inside the turnaround region turn inside out in redshift space in an ever more elongated fashion the closer they are to the core, producing the characteristic curved velocity caustic that bounds the artifact.  Thus, if the infall region of a cluster has the two-trumpet-horn shape, a fit SIM envelope is useful for estimating turnaround size, number of cluster members, and other things.

The standard method for fitting an envelope, introduced by RG89 and used by all authors since, is to first convolve the data about the line of sight cluster axis (i.e. plot radial velocity versus angular separation from the cluster centre). We will call this kind of redshift space {\bf projection space} or $\mbox{S}_p$, to distinguish it from {\bf slice space} or $\mbox{S}_v$.  Slice space is the usual redshift pie plot (i.e.  radial velocity versus some orientation with respect to the cluster centre such as right ascension, declination, or any other direction).

Studies find many clusters are not well fitted  by a SIM envelope  in $\mbox{S}_p$ (see e.g. RG89, Vedel \& Hartwick 1998,  Rines et al. 2003, AA11). Moreover, N-body simulations of flat universes demonstrated that the velocity fields surrounding clusters in $\mbox{S}_p$ often differ considerably from the predictions of SIM both because of the presence of substructure and recent mergers (van Haarlem 1992; van Haarlem \& van de Weygaert 1993; Diaferio \& Geller 1997) and because the shape of the velocity field changes when the line of sight is changed, making it difficult to judge the correctness of the predictions of SIM with respect to the velocity field.  This has led many to abandon SIM and turn to alternatives such as caustic technique (Diaferio 1999; Rines et al. 2003)

In this paper we offer another alternative for analyzing clusters:  tilted SIM.  By tilted, we mean SIM generalized to include flows that cause the infall artifact to tilt and no longer be axial symmetric.   The reason this could be a potentially useful tool is that some clusters which do not have a curving caustic profile in axially convolved projection space $\mbox{S}_p$ do in fact have such a profile in slice space $\mbox{S}_v$, but {\em tilted}; and clear only when the slice direction matches the direction of the tilt.  Therefore fitting a tilted SIM envelope in $\mbox{S}_v$ to such clusters is useful not just for more accurately estimating turnaround and cluster members but also for estimating the magnitudes of possible peculiar velocity flows. 

Two types of flow that can cause tilt are transverse motion between the observer and cluster,  and rotational motion (flow with curl) around the cluster center.  
We provide the equations for finding the tilted 2D envelope that results when any SIM is generalized to include both these types of flow, as well as for the tilted 3D envelope produced by SIM plus transverse motion.  We also introduce a new algorithm for fitting an envelope to data that is based on the number density of galaxies, and is somewhat similar to an earlier method described by Van Haarlem et al. (1993).

The paper is organized as follows. In \S\ref{sec:Red}, we illustrate the way redshift space artifacts produced by spherical infall plus transverse motion or rotational flow will tip.  In \S\ref{sec:infReg}, we present our technique for identifying infall regions, illustrating and testing with SIM-based toy model simulations.
In  \S~\ref{sec:Clusters} we illustrate the method by fitting tilted SIM envelopes to three real clusters. The cluster fits are discussed in \S\ref{sec:Disc}, and we finish in \S\ref{sec:Con} with a summary of the results and conclusions. Throughout the paper we select $\Omega_0=0.27$, and $h_0=0.73$ km s$^{-1}$ Mpc$^{-1}$ (see e.g. Freedman \& Madore 2010) and consider the local universe with redshift $z \lesssim 0.1$.

\section{The Redshift Space Infall Artifact} \label{sec:Red}

The observed velocity of a galaxy, of radial infall speed $s_{\text{rad}}$ and azimuthal angle $\phi$, on a shell of radius $r'$ centred on the cluster is given by
\begin{equation}		\label{eq:PSE} \begin{split}
s_{\text{obs}} &= (H_0 R - v_{0x})\cos\phi - \left(s_{\text{rot}}(r') \frac{R}{r'} + v_{0y}\right)\sin \phi   \\
	&  \qquad  \pm \left(H_0 r' - s_{\text{rad}}(r')\right) \left(1 - \left(\frac{R}{r'}\sin\phi\right)^2\right)^{1/2},
\end{split}\end{equation}
where $v_{0x}$ and $v_{0y}$ are the radial and transverse peculiar velocities of the observer, respectively, and $s_{\text{rot}}$ is rotational speed about the cluster centre (see Appendix~\ref{app:2Dsh} \& PS94). Notice that Eq.\ref{eq:PSE} is a generalized case of Eq.~(23) in RG89, which ignores the spatial velocity of the observer with respect to the cluster centre and assumes the flow is purely radial.

\subsection{Shells and envelope in redshift space}

SIMs have been extensively described in the literature (e.g. Gunn \& Gott 1972; Silk 1974; Gunn 1978; Peebles 1976; Lilje \& Lahav 1991). For illustrative purposes, the SIM we are using is the Praton-Schneider model (PSM) because it is conveniently parametrized in terms of the cluster virial speed and turnaround radius, making it easy to fit to the cluster observables virial velocity dispersion $\sigma_{\text{vir}}$ and angular turnaround radius $\alpha_{\text{turn}}$ (see PS94).  However, any SIM can be used without disturbing the general conclusions.   

\subsubsection{The 2D \& 3D Envelopes}

\begin{figure} \centering
\vspace{-0.2cm}
\includegraphics[width=25cm] {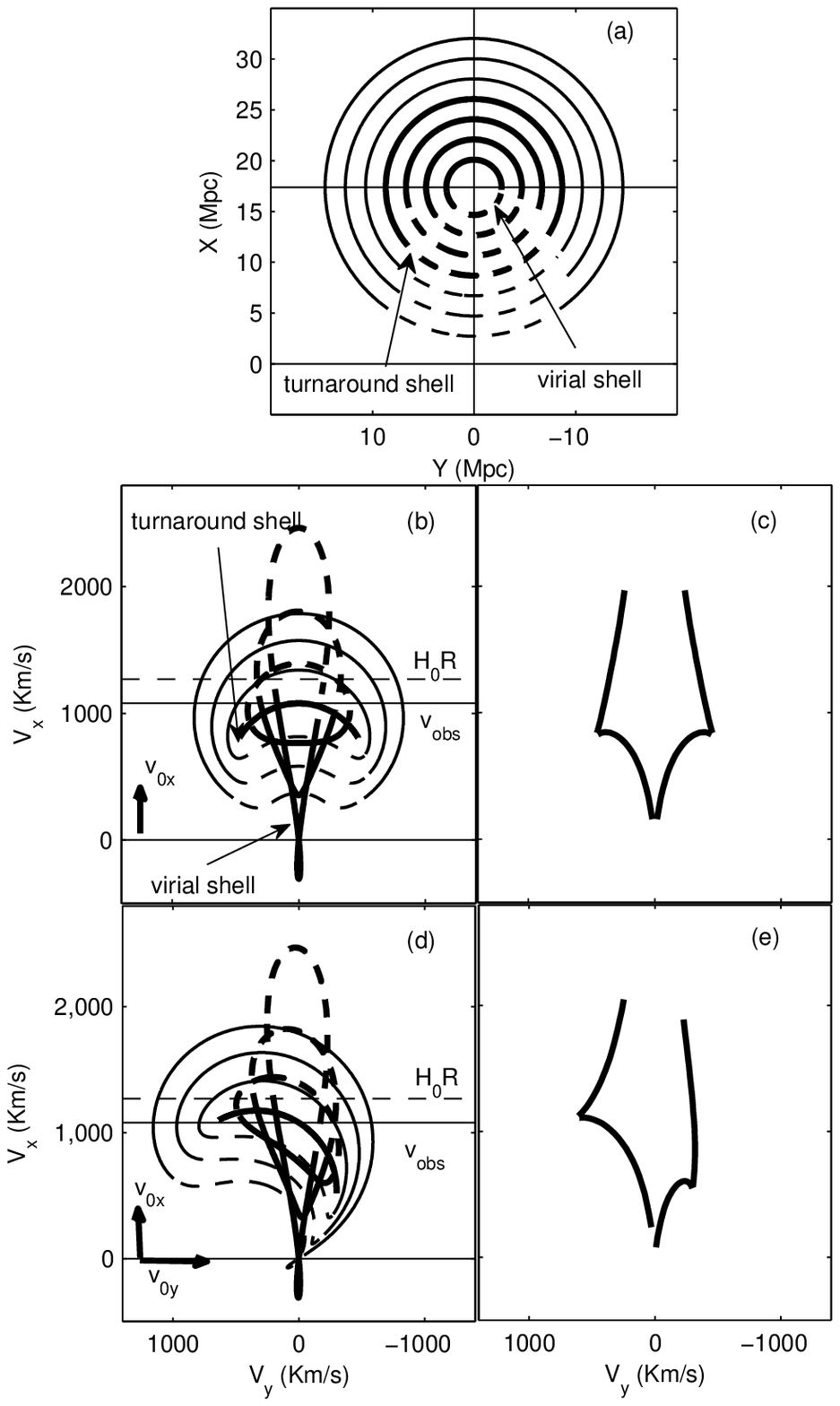} \vspace{-0.7cm}
\caption{Nested spherical shells in the field of a simulated cluster. Panel (a) shows the shells in real space. Panels (b) and (d) show the shells in the redshift spaces for $v_{0y}= 0$ and -670 km s$^{-1}$, respectively. Panels (c) and (e) show the application of the 2D envelope for the shells in panels (b) and (d), respectively.} 
\label{fig:shells2d}
\end{figure}

An example of the application of Eq.~\ref{eq:PSE} when there is no rotation ($s_{\text{rot}} = 0$) is shown in Fig.~\ref{fig:shells2d}. It shows a set of nested shells of a simulated cluster obeying PSM of the parameters $\alpha_{\text{turn}}=30^{\circ}$, $\sigma_{\text{vir}}=800$ km s$^{-1}$, $v_{\text{obs}}=1079$ km s$^{-1}$, R = 17.4 Mpc and $v_{0x}=\mbox{H}_0\mbox{R}-v_{\text{obs}}=190$ km s$^{-1}$, where $\sigma_{\text{vir}}$ is the velocity dispersion of the cluster at the virial radius.

The shells in Fig.~\ref{fig:shells2d} are shown in both real space ({\em up}) and redshift space ({\em down}), in the absence ({\em top panel}) and presence ({\em bottom panels}) of transverse velocity $v_{0y} = -670$ km s$^{-1}$. The velocity caustic envelope that bounds the cross section of the infall artifact in each case is also shown ({\em far right}). 
This figure is similar to unpublished ones in Praton (1993), hereafter P93; and the pair of panels showing the shells in the absence of transverse motion (Figs.~\ref{fig:shells2d}a \& b) is similar to a pair in Hamilton (1998).

In Fig.~\ref{fig:shells2d}, shells drawn with thin lines are outside the cluster's turnaround shell ({\em bold}) and shells drawn with thick lines are inside. The innermost shell represents the virial radius. The near side of each shell is dashed and the far side is solid line. Note that the shells inside the turnaround turn inside-out so that near side and far side reverse, while the turnaround shell collapses so that its near side and far side coincide to form a circular arc in this cross section view.  

The shells immediately outside the turnaround do not turn inside out, but are crowded together, so that some material outside the turnaround lies inside the envelope.  This illustrates the triple-value problem (see Tonry \& Davis 1981), where there are some foreground and background galaxies that appear to be part of the cluster because of the distortion in redshift space.

In the 3D redshift space $\mbox{S}_v$, a galaxy cluster looks like two trumpet horns glued together. Assuming spherically symmetric infall only and no rotational flow, the line of sight velocity $s_{obs}$ of a shell of radius $r'$ in 3D is given by 
\begin{equation} \label{eq:PSE3D} \begin{split} 
s_{\text{obs}}(\alpha, \beta) = (H_0 R - v_{0x}) \cos\alpha - v_{0y} \sin\alpha \cos\beta \pm \\ \qquad
(H_0 r' - s_{\text{rad}}(r')) \sqrt{1-\left(\frac{R}{r'}\right)^2 \sin^2\alpha},
\end{split}\end{equation}
\noindent where $\alpha$ is the polar angle that runs from 0 to $180^\circ$ and $\beta$ is the azimuthal angle that runs from 0 to $360^\circ$, defined relative to the cluster axis (see Appendix \ref{app:3DEnv}). Note that Eq.~\ref{eq:PSE3D} is exactly the same as Eq.~\ref{eq:PSE} that describes the 2D cross section of the shell in the x-y plane when $s_{\text{rot}}=0$. The only difference, besides changing $\phi$ by $\alpha$, is the additional factor of $\cos\beta$ multiplying the transverse velocity $v_{0y}$ term. 

\begin{figure}
\vspace{-0.2cm}
\includegraphics[width=26cm] {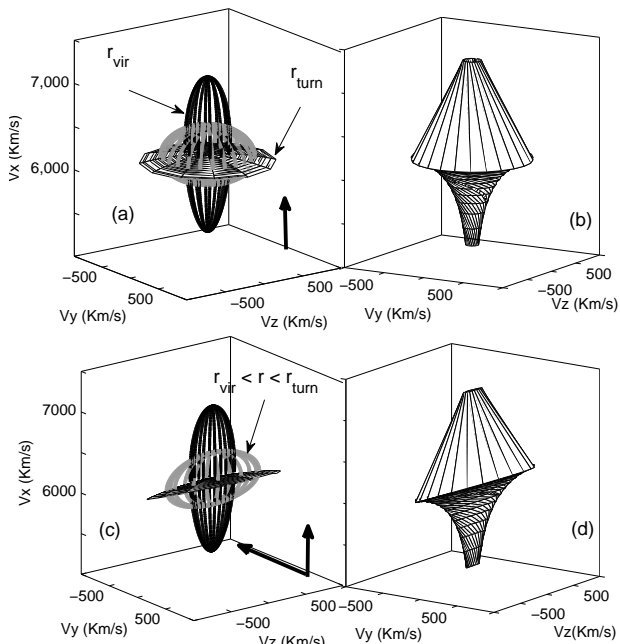} \vspace{-3.9cm}
\caption{Nested spherical shells in the field of a simulated cluster. Panels (a) and (c) show the shells in three dimensional redshift spaces for $v_{0y}= 0$ and -2640 km s$^{-1}$, respectively. Panels (b) and (d) show the three dimensional envelopes for the shells in panel (a) and panel (c), respectively.} 
\label{fig:shells3d}
\end{figure}

Fig.~\ref{fig:shells3d} shows the application of Eq.~\ref{eq:PSE3D} for 3D shells {\em (left)} and Eqs.~\ref{eq:3DenvAlpha} \& \ref{eq:3DenvSobs} for 3D envelopes {\em (right)}. The parameters of the envelope are $\alpha_{\text{turn}}=7^{\circ}$, $v_{\text{obs}}=6250$ km s$^{-1}$ and $\sigma_{\text{vir}} =516$ km s$^{-1}$. Note that the near and far side of the turnaround shell coincide to form a cupped dish in 3D redshift space, like a ring on the finger artifact (PS94). Figs.~\ref{fig:shells3d}b \& ~\ref{fig:shells3d}d show the 3D envelopes for the shells of Figs.~\ref{fig:shells3d}a \& ~\ref{fig:shells3d}c, respectively. As shown, the 3D envelope forms a two-trumpet-horn shape which is not tilted in Fig.~\ref{fig:shells3d}b and tilted in Fig.~\ref{fig:shells3d}d. This is the first attempt to obtain such a shape which was introduced by Kaiser (1987).

\subsection{Tilt}

Tilt caused by the observer's transverse velocity relative to the cluster points down in the direction the observer is going relative to the cluster. For example, in Fig.~\ref{fig:shells2d}d, the observer has transverse velocity to the right, relative to the cluster. In the observer's frame this is equivalent to all the shells having a transverse velocity to the left and causes the shells to tilt, down on the right and up on the left.  

Rotational flow about the cluster centre also produces tilt, down on the side moving towards the observer and up on the side moving away. For example, if the observer has no transverse velocity but the turnaround shell in Fig.~\ref{fig:shells2d}a was rotating clockwise about the cluster centre, in redshift space it would still tilt down on the right and up on the left.  

In fact, Eq.~\ref{eq:PSE} shows that if each shell has a rotational flow with magnitude $s_{\text{rot}} = (constant) \times r'$ in addition to its infall velocity, the result is the same as that produced by $v_{0y}$. In other words, infall plus transverse velocity is indistinguishable from infall plus solid body rotation.  Fig.~\ref{fig:shells2d}d would look identical in either case (P93, PS94).

Although we don't consider infall plus solid body rotation to be a likely scenario,  we point out that the possibility of flow with curl should be kept in mind when looking at tilted infall artifacts.  Exploration of various other toy models incorporating rotational flow shows they produce tilted structures often reminiscent of shapes seen in redshift surveys (P93).

The slope of the tilt (the  slope of line drawn from left-side point to right-side point of the envelope enclosing the artifact) is
\beq
\text{\it slope} = \left(v_{\text{circ}} \frac{R}{r_{\text{turn}}} + v_{0y}\right)/v_{\text{obs}}
\label{eq:slope}
\eeq
where $v_{\text{circ}} \equiv s_{\text{rot}}(r'=r_{\text{turn}})$ is the rotational speed (if any) of the turnaround shell.  The artifact tips up on the right if  observer is moving to left ($v_{0y} > 0$) or rotation is counterclockwise ($v_{\text{circ}} > 0$) and tips up on the left if observer is moving right ($v_{0y} < 0$) or rotation is clockwise ($v_{\text{circ}} < 0$). This is easily derived by considering the cross section of the turnaround shell in redshift space, since the ends of the arc shape coincide with the points of the envelope (see PS94 for figures and derivation).

We find slope is insensitive to the details of the envelope used to find it.  In practice, this means  slope can be determined from Eq.~\ref{eq:slope} and the best fit values of  $v_{obs}$ and $v_{0y}$ of an envelope generated by any SIM with no rotational flow.  Once the slope is known, it can be used to estimate the circular velocity needed to produce an equivalent tip, via Eq.~\ref{eq:slope} and the relation $r_{turn}/R = \sin(\alpha_{turn})$. In other words, use the following pair of relations to analyze a cluster:
\begin{equation}
\begin{array}{llcl}
v_{0y} & = slope \cdot v_{obs}	& \text{ if } &  v_{circ} = 0, \\
v_{circ} & = slope \cdot v_{obs} \sin \alpha_{turn}   & \text{ if } & v_{0y} = 0.
\end{array}
\label{eq:v0yvcirc}
\end{equation}

Note that the observed speed of the cluster is approximately its Hubble velocity ($v_{\text{obs}} \sim H_0 R$).  So, if the tilt of the artifact is due {\em only} to transverse velocity $v_{0y}$  of the observer relative to the cluster or due {\em only} to rotational flow $v_{circ}$  of the turnaround region , then the flow responsible has approximate magnitude
$v_{0y} \sim (\text{\it slope}) \times H_0 R $ (if $v_{circ} = 0$) or $v_{circ} \sim (\text{\it slope})\times  H_0 r_{\text{turn}}$ (if $v_{0y} = 0$).

So, if there are no rotational flows around clusters we expect that the farther away the cluster the smaller the tilt, in general.  However, if there are rotational flows or, more generally, flows with curl, we don't expect to see a decrease in tilt with distance.

\subsection{Width-to-Length Ratio}\label{sec:wtl}

\subsubsection{Width Variation in Redshift Space}
\label{sec:WidVarInRedSpace}

\begin{figure}
   \centering \vspace{-0.3cm}
   \includegraphics[width=7cm]{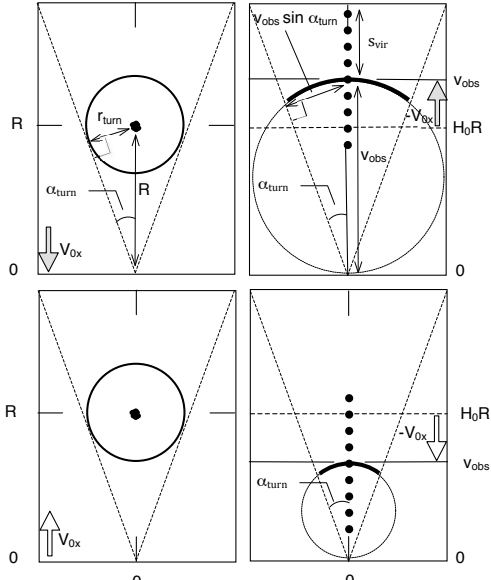} 
   \vspace{-0.2in}
   \caption{\small Turnaround region in real space ({\em left}) and redshift space ({\em right}) when observer's velocity $v_{0x}$ is away from the cluster ({\em top}) and towards the cluster ({\em bottom}).  In each case,  the linear width of the redshift space artifact is $v_{\text{obs}} \sin\alpha_{\text{turn}}$, where $v_{\text{obs}} = H_0 R - v_{0x}$ is the observed velocity of the cluster.}
   \label{fig:posnegV0x}
\end{figure}

Consider the redshift space artifact produced by the turnaround region of radius $r_{\text{turn}}$ surrounding a galaxy cluster, in the special case where the observer at the origin has no radial peculiar velocity with respect to the cluster, which lies a distance $R$ away along the $x$-axis (Fig.~\ref{fig:posnegV0x}). The width of the artifact measured by the observer (the distance from the outer point to the central axis) will be its intrinsic width:  $H_0\, r_{\text{turn}}$, where $H_0$ is the Hubble constant. 

However, if the observer has non-zero radial peculiar velocity $v_{0x}$ (Fig.~\ref{fig:posnegV0x}),  then in redshift space the cluster's position will shift away from the observer if the observer's velocity is away from the cluster ($v_{0x} < 0$), and the cluster's position will shift towards the observer if  the observer's velocity is towards the cluster ($v_{0x} > 0$).
The angular size of the cluster will not change, since this depends on the size of the turnaround region $r_{\text{turn}}$ and the distance to the cluster R, and $r_{\text{turn}}$ and R do not change if the observer has peculiar velocity. But, the linear width {\em does} change: it is more wide if $v_{0x} < 0$ (cluster farther in redshift space than real space)  and less wide if $v_{0x} > 0$ (cluster closer in redshift space than real space).

The artifact's length (the length of the central finger) of course will not change, since this depends only on the mass and radius of the cluster's core.  For convenience, we will take this length to be the virial speed $s_{\text{vir}} = \sqrt{3} \sigma_{\text{vir}}$ where $\sigma_{\text{vir}}$ is the observed velocity dispersion of the cluster at virial radius.

So, if the observer has non-zero radial velocity, the apparent ratio of the artifact's width to length  will change from its intrinsic value.  If $v_{0x} > 0$ the ratio will decrease (artifact is less wide), and if $v_{0x} < 0$ the ratio will increase (artifact is more wide).

The apparent width of the artifact is $v_{\text{obs}} \sin(\alpha_{\text{turn}})$, so to find the ratio $\mathcal{W}$ of apparent width to length $s_{\text{vir}}$ when analyzing a cluster, use the following relation:
\begin{equation}
\mathcal{W} = \frac{v_{\text{obs}} \sin(\alpha_{\text{turn}}) }{ \sqrt{3} \,\sigma_{\text{vir}} }.
\label{eq:wtl}
\end{equation}

In the expression above, $v_{\text{obs}} = H_0 R - v_{0x}$ is the observed speed of the cluster core, $R$ is the distance to the cluster, and $\alpha_{\text{turn}}$ is the observed angular width of the artifact, with $\sin\alpha_{\text{turn}} = r_{\text{turn}}/R$.  So, Eq.~\ref{eq:wtl} can be rewritten 
\begin{equation}
\mathcal{W} = \frac{v_{\text{obs}} \, r_{\text{turn}}/R}{s_{\text{vir}}} = \frac{(H_0 \,R - v_{0x})  \,r_{\text{turn}}/R}{s_{\text{vir}}}.  
\end{equation}
Note that if $v_{0x} = 0$, then $v_{\text{obs}} = H_0 R$ and the ratio of width to length becomes equal to the intrinsic width to length ratio $\mathcal{W}_0$:  i.e., $\mathcal{W} = \mathcal{W}_0 \equiv H_0 \,r_{\text{turn}} / s_{\text{vir}}$, as expected.

We can write the observed ratio $\mathcal{W}$ as a sum of the intrinsic ratio $\mathcal{W}_0$ plus an extra term:  
$\mathcal{W} = \mathcal{W}_0 + \Delta \mathcal{W}$, where 
\begin{equation}
\frac{\Delta \mathcal{W}}{\mathcal{W}_0} = -\frac{v_{0x}}{H_0 R}.
\label{eq:deltaWW0}
\end{equation}  
 
Once the observer's velocity is known, the distance $R$ to the cluster can be found using 
 \begin{equation}
 R = (v_{\text{obs}} + v_{0x})/H_0
 \label{eq:H0R}
 \end{equation}

The distortion in width to length caused by the observer's radial motion is thus inversely dependent on distance (just as tilt caused by observer's transverse motion is inversely dependent on distance). In other words, the farther away the cluster lies, the larger $v_{0x}$ must be to produce the same amount of change in the width to length ratio $\mathcal{W}$.

\subsubsection{Determining Observer's Velocity Towards a Cluster}

Suppose we know what the {\em intrinsic} width to length ratio $\mathcal{W}_0$ of the infall region of a given cluster ought to be.  If that is the case, then we can use the observed width to length ratio $\mathcal{W}$ to determine the observer's radial velocity $v_{0x}$ with respect to the cluster, as follows.
 
The observed velocity of the cluster is $v_{\text{obs}} = H_0 R - v_{0x}$.  Combining this with Eq.~\ref{eq:deltaWW0}  and solving for $v_{0x}$ yields
 \begin{equation}
 v_{0x} = -\left(\frac{\Delta\mathcal{W}}{\mathcal{W}}\right) v_{obs} =  \left(\frac{\mathcal{W}_0}{\mathcal{W}} - 1\right) v_{\text{obs}},
 \label{eq:v0x}
 \end{equation}
 where a positive value means observer has peculiar velocity towards the cluster and a negative means motion away.
 The radial velocity of the observer relative to the cluster can thus be determined using only the observed velocity of the cluster and the observed difference in the infall artifact width to length ratio compared to what's expected. 

In practice, most clusters are distant enough that we don't expect $\mathcal{W}$ to vary much from whatever the intrinsic ratio $\mathcal{W}_0$ is for that particular cluster.  For example, if observer's radial peculiar velocity with respect to a cluster was as large as our motion with respect to the cosmic microwave background ($v_{0x} \approx 600 \mbox{ km s}^{-1}$), then the change in the width to length ratio of an infall artifact of a cluster at the redshift of  Coma ($v_{\text{obs}} \approx 6000 \mbox{ km s}^{-1}$) is only about 10\% ($\Delta\mathcal{W}/\mathcal{W}_0 = -0.09$) and the change in the ratio for more distant clusters would be even less.
 
However, the technique could be useful for a nearby cluster such as Virgo.   For example, suppose it turns out that the infall artifacts of all clusters have the same $\mathcal{W}_0$; or that $\mathcal{W}_0$ has a direct  and predictable relationship to the observed density distribution.  Then one could determine $\mathcal{W}_0$ for Virgo by doing a study of distant clusters (for which $\mathcal{W} \approx \mathcal{W}_0$), then use that value in Eq.~\ref{eq:v0x} to make an estimate of our relative radial peculiar velocity with respect to Virgo.  Note that, unlike methods based on distance estimates, this method of estimating the relative radial peculiar velocity is independent of the value of $H_0$.

\section{Identifying Infall Regions}\label{sec:infReg}

In this section we describe our method for fitting envelopes to tilted infall regions.   
To illustrate and test the method, we employ spherical infall toy models (see Kaiser 1987; RG89; P94), rather than n-body. We follow PS94 to construct the simulation. We use the toy SIM because clusters in the publicly available outputs of $\lambda$CDM cosmological simulations we have investigated so far, such as GIF project (Kauffman et al. 1999), Bolshoi simulation (Klypin, Trujillo-Gomez, \& Primack 2011), and Multi-Dark Run 1 (Prada et al. 2012), have little infall distortion and no tilt.  Thus we cannot use existing n-body simulation to test whether our fitting method correctly recovers peculiar velocities that cause such tilt.
\begin{figure} 
\vspace{-0.2cm}
\includegraphics[width=23cm] {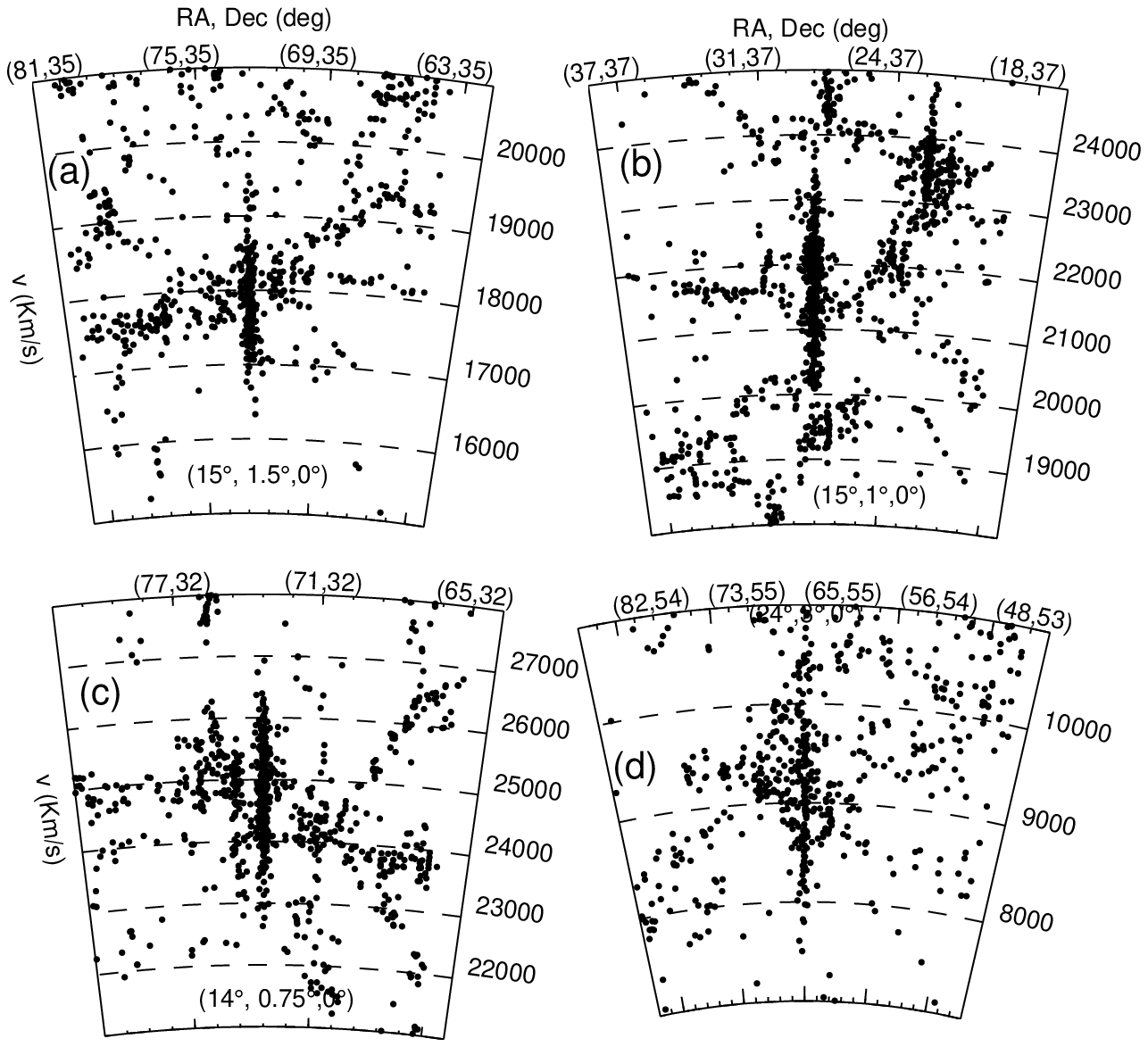} \vspace{-3.1cm}
\caption{The distributions of galaxies of an n-body simulation (Yoshida, Sheth \& Diaferio 2001) in the field of four galaxy clusters.} 
\label{fig:simcls}
\end{figure}

As an example of this problem, Fig.~\ref{fig:simcls} 
shows a sample of four clusters from about 20 investigated clusters taken from an n-body simulation with $\Omega_0= 0.3$, $\Omega_{\Lambda} = 0.7$ and $h_0=0.7$ km s$^{-1}$ Mpc$^{-1}$  (Yoshida, Sheth \& Diaferio 2001).  The thickness of each slice is chosen to satisfy an optimal thickness of 2/3 $\alpha_{\text{turn}}$ (see \S\ref{sec:Clusters}), supposing $r_{\text{turn}}= 4 r_{\text{vir}}$, and each cluster was investigated in a full range of orientations.  

\begin{figure}
   \centering
\vspace{-0.4cm}
  \includegraphics[width=26cm]{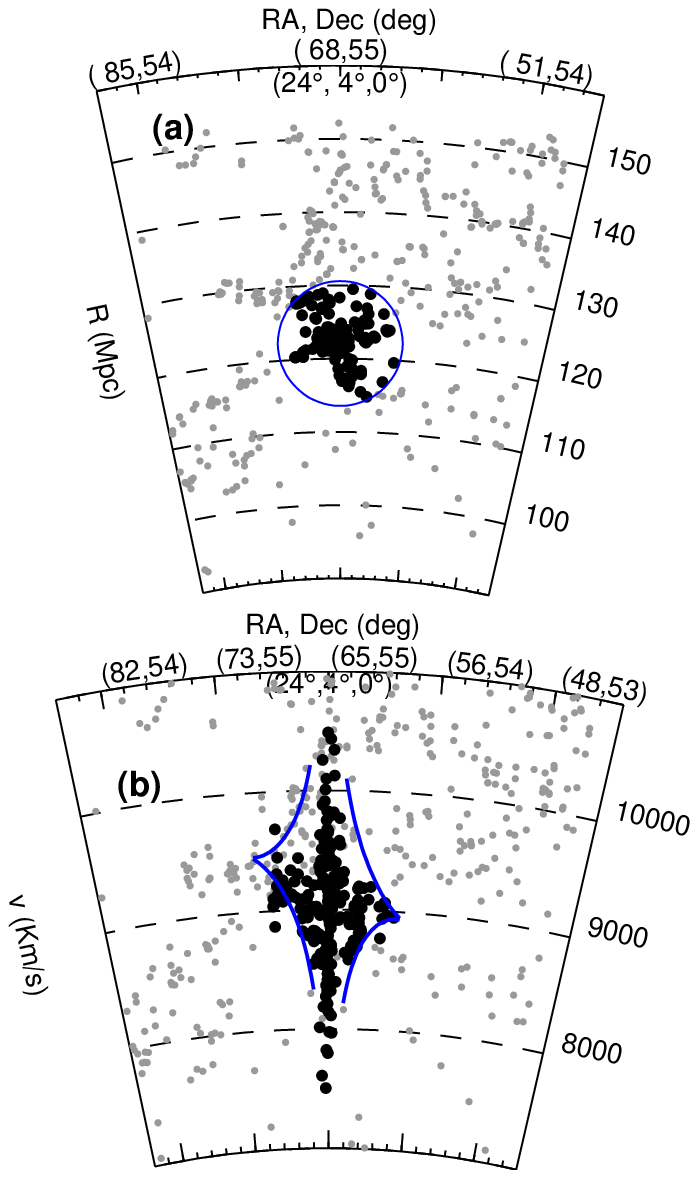} \vspace{-3.4cm}
   \caption{`Tilted' n-body cluster (Fig~\ref{fig:simcls}d) in (a) real space and (b) redshift space, with SIM fit ({\em envelope}) and corresponding turnaround radius ({\em circle}) indicated.  Black points are galaxies inside circle.}
   \label{fig:simcls1}
\end{figure}

Most of the 20 clusters have little obvious infall distortion (like those in Fig.~\ref{fig:simcls}a--c), but a single cluster (panel d) has a structure that resembles the tilted infall-artifact-like shapes seen in survey data (see \S\ref{sec:Clusters}).  However, this turns out to be mostly real structure, and not infall distortion
(Fig.~\ref{fig:simcls1}).
Note the galaxies inside the turnaround assumed for an envelope fit to the apparent artifact do not fill the envelope.  In particular, note the curving edge of the top left of the artifact-like structure is actually a filament lying well away from the cluster; and the true turnaround is smaller than that assumed for the fit envelope.  This is different from the Virgo cluster (Fig.~\ref{fig:Virgo1}), and Virgo also has a different peculiar velocity field (Fig.~\ref{fig:VirNbodSim}). 

\subsection{Comparison between $\mbox{S}_p$ and $\mbox{S}_v$} \label{sec:SpSv}

\begin{figure} 
\vspace{-0.3cm}
\includegraphics[width=25cm] {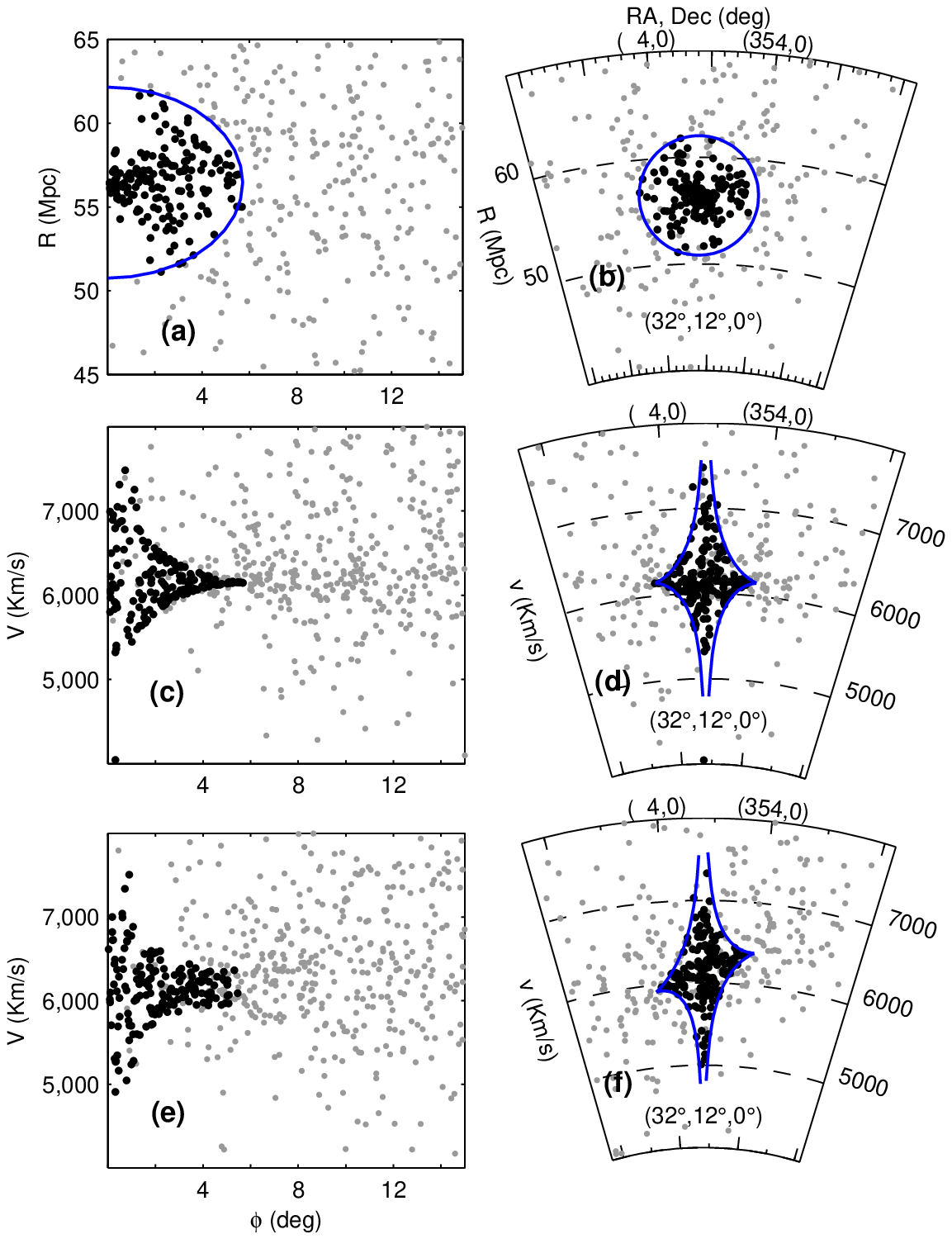} \vspace{-1.4cm}
\caption{Projection space $\mbox{S}_p$ (left column) compared with slice space $\mbox{S}_v$ (right column) for SIM-based simulated cluster. Slice thickness is $12^{\circ}$.  Panels (a, b) shows real space. Circle shows turnaround radius. Panels (c, d) show redshift space, with no transverse motion. Panels (e, f)  show redshift space, with transverse motion ($v_{0y} = -2465$ km/s). Curved lines in panels d \& f show SIM envelope.}
\label{fig:RealSpSv}
\end{figure}

Fig.~\ref{fig:RealSpSv} shows the distribution of galaxies of a simulated cluster with velocities obeying PSM, in real and redshift spaces. The simulation parameters are set such that angular size of the turnaround $\alpha_{\text{turn}}=5.7^\circ$, viral dispersion $\sigma_{\text{vir}}=516$ km s$^{-1}$ , observed speed $v_{\text{obs}}=6180$ km s$^{-1}$, distance R = 56.4 Mpc and observer's velocity toward the cluster $v_{0x}=-2061$ km s$^{-1}$.

Fig~\ref{fig:RealSpSv}a shows real $\mbox{S}_p$, while Figs.~\ref{fig:RealSpSv}c \& ~\ref{fig:RealSpSv}e show redshift $\mbox{S}_p$, where a galaxy's observable velocity is plotted versus its angular separation from the cluster centre (see e.g., RG89; Rines et al. 2003; AA11). Figs.~\ref{fig:RealSpSv}d \& ~\ref{fig:RealSpSv}f show two slices for the distribution of the galaxies in $\mbox{S}_v$ where the observable velocity is plotted versus right ascension.
Note that in redshift space (panels c - f) the cluster is elongated in the core region and compressed further out, compared to its distribution in real space (panels a \& b).

In Figs.~\ref{fig:RealSpSv}c \& ~\ref{fig:RealSpSv}d we compare the appearance of the simulated cluster in $\mbox{S}_p$ and $\mbox{S}_v$, respectively, when no transverse or rotational velocities are present. As shown, the caustic-shape pattern of the cluster appears clearly in both $\mbox{S}_p$ and $\mbox{S}_v$ and SIM can successfully describe this pattern in either space.  Panels (e) and (f) show what happens when the observer has a transverse velocity  $v_{0y}=-2465$ km s$^{-1}$. In $\mbox{S}_p$ the caustic-shape pattern does not appear clearly and according to van Haarlem \& van de Weygaert (1993) and Diaferio \& Geller (1997), this is because of the existence of random motion in the outer region of the cluster and therefore SIM fails to describe the cluster. But in this particular case there is a transverse motion between the cluster and the observer causing a tilt in the structure, and tilts are obscured in $\mbox{S}_p$. By contrast, the tilt appears clearly in $\mbox{S}_v$ and SIM can describe this pattern successfully. Therefore, not all the galaxies, particularly at the outer region of the cluster, are members. Some of them are members and the others are not, although they all seem to be settled in the cluster. 

\begin{figure} 
\vspace{-0.5cm}
\includegraphics[width=20cm] {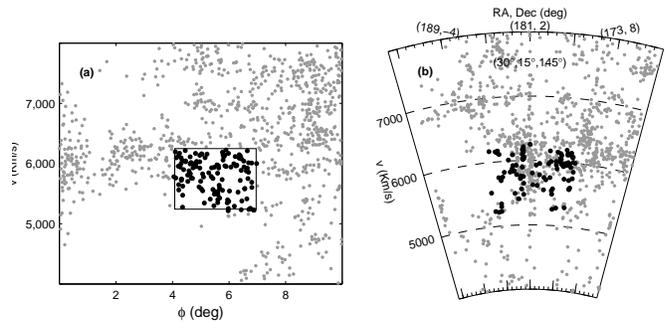} \vspace{-5.2cm}
\caption{A comparison between the distribution of galaxies in $\mbox{S}_p$ (panel a) and $\mbox{S}_v$ (panel b) of A1459.} 
\label{fig:Projection}
\end{figure}

Accordingly, the disadvantages of projection space $\mbox{S}_p$ are as follow. (1) First, transverse velocity (if any) of the observer with respect  to the cluster and  rotational flow (if any) about the cluster's centre do not appear in $\mbox{S}_p$.  In other words, if such motions are present, it is impossible to tell because the data are convolved. (2) Second, the calculation of the angular separation depends on the choice of the cluster centre and any change in this choice will change the appearance of the caustic shape and make it difficult to determine the true turnaround radius. In other words, the shape of the cluster changes with changing the line of sight with respect to the cluster centre (van Haarlem \& van de Weygaert 1993). (3) Third, calculating the angular separation does not take into account if a galaxy is on the right or the left of the cluster centre and this may cause a critical situation, where one can see a group of galaxies in the cluster field and consider them as a real group, although some of them may be on the right and the others are on the left and they are very far from each other. This effect is shown in Fig. ~\ref{fig:Projection} for A1459 (see \S\ref{sec:Clusters}), where $\mbox{S}_p$ is plotted in panel (a) and $\mbox{S}_v$ is plotted in panel (b). As shown, the black galaxies within the rectangular seem to create a substructure but when plotted in $\mbox{S}_v$ these galaxies take their correct positions and do not show a true substructure.

\begin{figure*} 
\vspace{-0.5cm}
\includegraphics[width=18cm] {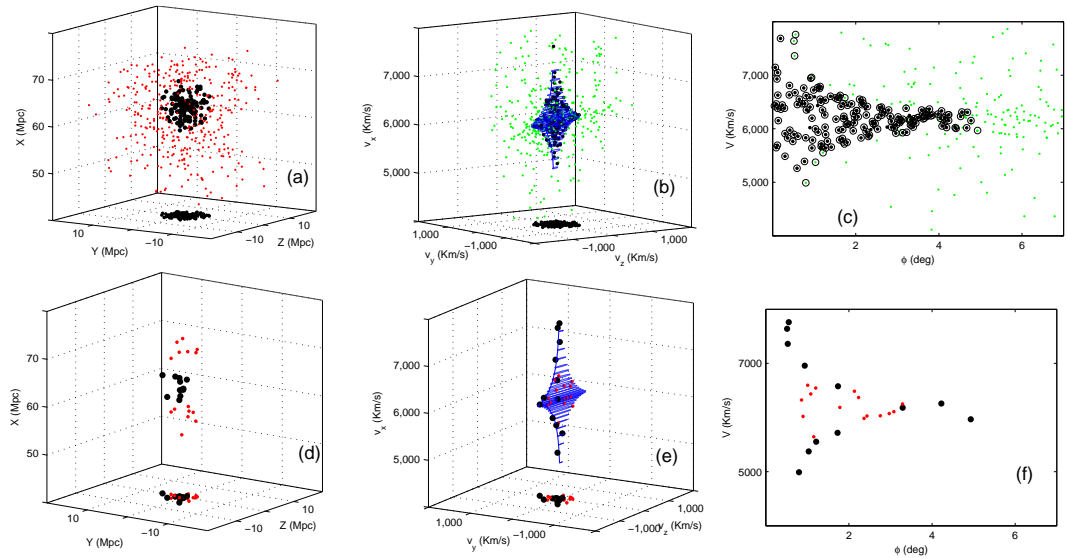} \vspace{-1.1cm}
\caption{A simulated galaxy cluster in 3D. Panel (a) shows 3D real space of the cluster. Panel (b) shows 3D redshift space, where the big points indicate the members of the 3D envelope and the small points are the outliers located outside the 3D envelope. Only the true members are projected on y-z plane. Panel (c) shows $\mbox{S}_p$, where the big and small points are the same as panel (b) and circles are the 171 true members. Panels (d-f) are the same as panels (a-c), respectively, but for the true member located outside the envelope ({\em big points}) and true outliers settled within the 3D envelope ({\em small points}) only.}
\label{fig:3DSim}
\end{figure*}

By contrast, the advantages of slice space $\mbox{S}_v$ are as follows. (1) First, one can control the thickness of the slice. As the thickness increases, more galaxies will be projected on the slice. A thin slice is important for showing the true shape of the cluster, whether it is tilted or not, whether the caustic-shape pattern appears or not, and where the location of the boundary of the cluster is in order to obtain the cluster's turnaround radius. A thick slice shows a general view of the cluster with its surroundings to know whether there are interactions with neighbor clusters or groups or not. (2) Second, one can control the orientation of the slice.   In many cases the true features of the cluster may not appear clearly in a slice taken in some default orientation (such as along constant RA or Dec) but do appear in another orientation. (3) Finally, in $\mbox{S}_v$ we can explore the cluster in the three dimensions, which is important for determining the cluster members to overcome the problem of projection (see \S\ref{sec:MemSel}).
\subsection{Membership Selection} \label{sec:MemSel}

The methods developed to determine clusters members can be classified into two categories: algorithms that use only the redshift information, e.g. 3$\sigma$-clipping techniques (Yahil \& Vidal 1977), fixed gapping procedures (Beers et al. 1990, Zabludoff et al. 1990), and jackknife technique (Perea et al. 1990); and the methods that use information of both position and redshift, such as the shifting gap procedure designed by Fadda et al. (1996), the virial theorem method introduced by den Hartog \& Katgert (1996), the caustic technique developed by Diaferio (1999), and SIM fitted in projection space $\mbox{S}_p$ (AA11). 

Here, we explore another approach to identify cluster members: fitting SIM in 3D space rather than $\mbox{S}_p$. In \S\ref{sec:Red} we introduced the 3D envelope that describes this pattern.  If one fits a 3D envelope to a given cluster and then takes the galaxies inside the envelope, the result may be a close approximation of the set of true members (galaxies which lie within the turnaround radius).   This approach is tested for a simulated cluster 
(Fig.~\ref{fig:3DSim}) of the parameters $v_{\text{obs}}=6180$ km s$^{-1}$, $\sigma_{\text{vir}}=516$ km s$^{-1}$, $\alpha_{\text{turn}}=5.1^\circ$ and $v_{0y}=2260$ km s$^{-1}$ and 171 true members.

We determine the galaxies within this envelope ({\em big points}) and consider them as the cluster members. The galaxies outside this envelope ({\em small points}) are considered as outliers. The comparison between the true members and the members that have been determined by the 3D envelope is shown in Fig.~\ref{fig:3DSim}c in $\mbox{S}_p$. 
Note that there is a loss of few of the true members mostly within the inner core of the cluster in which the velocities of the galaxies are assumed to be distributed randomly and SIM is not valid within this inner core region ( see Figs. ~\ref{fig:3DSim} d-f). 
Also, there are some outliers located within the 3D envelope, which illustrates the triple-value problem that is discussed in \S\ref{sec:Red}. As shown, although the true outliers are clearly located outside the cluster in real space (panel d) they are falling inside the cluster in redshift space (panels e-f). 

Notice that the cluster's members are obtained from all the cluster field, not from galaxies within a slice. Therefore, the determined members are not affected by the thickness or orientation of some slice. The reason to do slicing is diagnostic:   correctly oriented slices make the true features of the cluster (such as tilt) more obvious and are useful for determining physical parameters of a fit.  That fit can then be used to obtain the cluster's members from all the field.
\subsection{Envelope Fitting: NDM} \label{sec:NDM}

Fitting infall models such as PSM, Reg\H{o}s-Geller model (RG89) or Yahil approximation (Yahil 1985) to the distribution of galaxies is subjective if done by eye, and different people may get quite different results.
Therefore, we introduce an objective method, called Number Density Method (hereafter NDM), to determine the parameters of a given model that best fit the distribution of galaxies.

NDM depends on the calculation of the number density inside and outside the envelope.  The greater the number density is inside and the less is outside, the better the fit is.
The method can be used in any redshift space, whether $\mbox{S}_p$ or $\mbox{S}_v$, and for any model. For example, for Yahil approximation in which the only free parameter is $\Omega_0$, applying NDM gives the $\Omega_0$ which gives the best fit with the distribution of galaxies.
For PSM, applying NDM yields the best fit $\alpha_{\text{turn}}$ and $v_{0y}$ (or $v_{\text{circ}}$).

\subsubsection{Description of the Numerical Density Method}

We explain the steps of the method in redshift space $\mbox{S}_v$. 
(i) Put a cutoff range, $v_{\text{gap}}$, in the radial velocity direction with respect to the radial velocity of the cluster $v_{\text{obs}}$. 
(ii) Determine the area $A$ enclosed by the envelope and $v_{\text{gap}}$. 
(iii) Determine the number of galaxies within the envelope, $N_{\text{plus}}$. These galaxies contribute positively because the model includes them. 
(iv) Now, what about the galaxies which are outside the envelope? These galaxies contribute negatively, so we do not want to include all the galaxies in the field.  Doing that may cause biased results because there may be other neighboring galaxy clusters or groups.  We only want galaxies that could be members of the cluster.  Accordingly, let  $N_{\text{minus}}$ be the number of galaxies that are located between the envelope under consideration and another larger envelope which has a turnaround radius greater than the first one by $r_b$.  This will be more obvious below. 
(v) Calculate the number density $N_{\text{den}}=(N_{\text{plus}} - N_{\text{minus}})/A$.

\begin{figure}\vspace{-0.4cm}
\includegraphics[width=14cm] {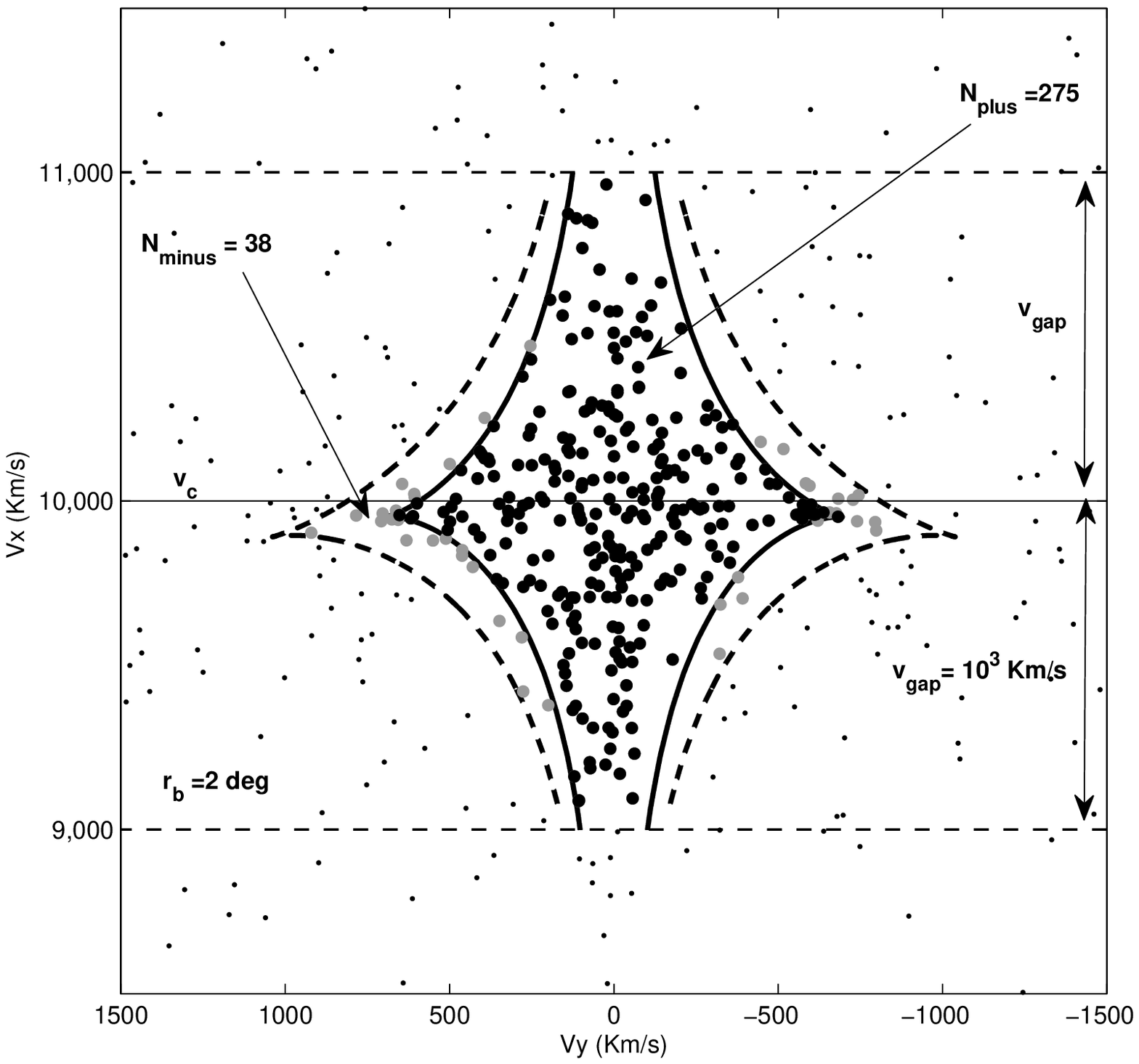} \vspace{-1.2cm}
\caption{Application of NDM. The black points within the solid envelope are the galaxies which contribute positively. The gray points indicate the galaxies which contribute negatively.}
\label{fig:Method}
\end{figure}

Fig.~\ref{fig:Method} gives an example of the application of NDM. The solid horizontal line indicates the location of $v_{\text{obs}}$ and the dashed horizontal lines indicate the location of $v_{\text{gap}}$ = 1000 km s$^{-1}$. The solid envelope has the parameters $\sigma_{\text{vir}}=620$ km s$^{-1}$, $v_{\text{obs}}=10000$ km s$^{-1}$, $\alpha_{\text{turn1}}=2^{\circ}$, for which we want to determine $\mbox{N}_{den}$. The area enclosed within this envelope is $1.1\times 10^6$ km$^2$ s$^{-2}$. The number of the galaxies (black points) within this envelope is $N_{\text{plus}} = 275$. The dashed envelope is the envelope within which we want to get the galaxies that contribute negatively. This envelope has $\alpha_{\text{turn2}}=4^{\circ}$ or $r_b=\alpha_{\text{turn2}}-\alpha_{\text{turn1}}=2^{\circ}$. The gray points are the galaxies which contribute negatively, then 
 $N_{\text{minus}} = 38$. Therefore, $N_{\text{den}} =(275-38)/(1.1\times 10^6)=2.2\times 10^{-4}$. Accordingly, the model with the parameters that give the highest value of 
 $N_{\text{den}}$ is considered as the best fitted model.

\subsubsection{Test NDM on Simulation}

To test the accuracy of NDM, we use it to fit PSM envelopes to PSM toy model simulated galaxy clusters (see e.g., RG 89, PS94).  Our goal is to make sure NDM can recover the input parameters in this special case. We test the method on $\alpha_{\text{turn}}$ and $v_{0y}$ with knowing the other parameters of the model. For 10 simulated clusters with different densities and at different distances we apply NDM to get the best $\alpha_{\text{turn}}$ and $v_{0y}$ for each simulated cluster (see Table \ref{tab:sim}). 

In general the application of NDM on simulation to determine the best envelope gives good results. In particular, the ratio $r_1 = (\alpha_{out} - \alpha_{in})/\alpha_{in}$, giving the percentage of error in determining $\alpha_{\text{turn}}$, gives an accuracy of nearly 97\%. Also, the ratio $r_2 = \sigma/\alpha_{in}$ shows that the standard deviation of the determined $\alpha_{\text{turn}}$ from NDM is, at maximum, less than 5\%. The ratio $r_3 = (v_{0y,out} - v_{0y,in})/v_{0y,in}$ gives an accuracy of about 96\% except for one simulated cluster for which the accuracy is 88\%. Finally, the standard deviation of the determined $v_{0y}$ from NDM, is less than 12.5\%.

NDM depends on the choice of $v_{\text{gap}}$ and $r_{b}$. Because galaxy clusters extend in redshift space in a range of about 2000 : 5000 km s$^{-1}$ and because spherical infall models can not be applied within the clusters' cores we suggest that $v_{\text{gap}}$ could range from about 700 : 1400 km s$^{-1}$ according to the extension of the studied cluster. The choice of $r_b$ should be selected carefully based on some factors because it may give improper results. The first one is the distance to the cluster from the observer, where $r_b$ should decrease with increasing the cluster's distance. For example, $r_b$ may range from 10 : 30 $^{\circ}$ for a cluster such as Virgo ($\approx$ 1079 km s$^{-1}$) and 5 : 15 $^{\circ}$ for a cluster such as A1459 ($\approx$ 6180 km s$^{-1}$). Second, $r_b$ should be large enough in the way that it is not affected by the inner density of the cluster, where the cluster's number density decreases with increasing distance from the cluster centre, so choosing $r_b$ large enough will remove the effect of decreasing density with increasing radius. On the other hand, selecting $r_b$ less than a certain value will give the best fit for parameters such as $\alpha_{\text{turn}}$ or $\Omega$ for smaller values. Also, $r_b$ shouldn't be larger than a certain value to avoid including galaxies from other groups or clusters.

\begin{table*} \centering
\caption{List of the parameters of 10 SIM-based simulated clusters with the application of NDM to get the goodness of the method.}	
\label{tab:sim}
\begin{tabular}{cccccccccccccccccc}\hline
\multicolumn{7}{c|}{Parameters' Model}&&\multicolumn{9}{c|}{Application of NDM}\\
\cline{1-7}\cline{9-17}
$v_{\text{obs}}$&$\sigma_{\text{vir}}$&R &$N_{sh}$&$\alpha_{in}$&$v_{0y,in}$&$v_{0x}$&&$\alpha_{out}$&$\sigma$&$r_1$&$r_2$&&$v_{0y,out}$&$\sigma$&$r_3$&$r_4$\\ 
km s$^{-1}$	&km s$^{-1}$ &Mpc	&	&($^\circ$)&km s$^{-1}$  &km s$^{-1}$ &&($^\circ$)&($^\circ$)&(\%)&(\%)&&	km s$^{-1}$ &km s$^{-1}$ &(\%)	&	(\%)\\
(1)		&(2) &(3)	&(4)	&(5)&(6)  &(7) &&(8) &(9) & (10)&(11)&&(12)	&(13) & (14) &(15)\\
\hline
1800		&757 &16.4	&5	&30.0&-450  &-600 &&30.5 &0.78 & 1.7&2.6&&-455	&34 & 1.1 &-7.6\\
3500		&660 &39.7	&5	&10.4& 1000 &-601 &&10.5 &0.18 & 1.0&1.7&& 957  &103&-4.3 & 10.3\\
6000		&783 &74.0	&4	&6.6 &-2500 &-599 &&6.51 &0.10 &-1.4&1.5&&-2505 &52 & 0.2 &-2.1\\
7500		&820 &94.6	&3	&5.4 & 3300 &-592 &&5.48 &0.14 & 1.5&2.6&& 3368 &196& 2.1 & 5.9\\
10000	  &785 &128.6	&3	&3.8 & 4000 &-609 &&3.86 &0.05 & 1.6&1.3&& 4086 &197& 2.2 & 4.9\\
12400	  &753 &161.6	&2	&2.9 &-3500 &-600 &&2.98 &0.09 & 2.8&3.1&&-3415 &187&-2.4 &-5.3\\
15000	  &824 &197.3	&3	&2.6 &-1550 &-599 &&2.64 &0.08 & 1.5&3.1&&-1728 &190& 11.5&-12.3\\
18000	  &728 &238.5	&2	&1.9 & 5000 &-592 &&1.88 &0.09 &-1.1&4.7&& 5102 &253& 2.0 & 5.1\\
20000	  &769 &265.9	&2	&1.8 & 6000 &-590 &&1.80 &0.05 & 0.0&2.8&& 6147 &212& 2.5 & 3.5\\
22400	  &720 &298.7	&2	&1.5 &-8300 &-594 &&1.55 &0.07 & 3.3&4.7&&-8240 &463&-0.7 &-5.6\\
\hline
\end{tabular}

\footnotesize{Cols 1-3: radial velocity, velocity dispersion and the distance to the cluster, respectively. Col 4: number of galaxies per shell. \\Col 5: turnaround angle. Cols 6-7: transverse and radial peculiar velocities of the observer, respectively. Cols 8-9: mean and standard deviation of $\alpha_{\text{turn}}$ determined by NDM, respectively. Cols 10-11: $r_1 = (\alpha_{out} - \alpha_{in})/\alpha_{in}$, $r_2 = \sigma/\alpha_{in}$. Cols 12-13: mean and standard deviation of $v_{0y}$ for the application of NDM, respectively. Cols 14-15: $r_3 = (v_{0y,out} - v_{0y,in})/v_{0y,in}$, $r_4 = \sigma/v_{0y,in}.$}
\end{table*}

One advantage of NDM  is that it can be applied to one, two, three or four quarters of the envelope, where in some cases the caustic shape of the distribution of the galaxies may not be clearly shown in one or more quarters of the cluster's field. Moreover, it is not affected by the presence of near groups or clusters of galaxies because we can control the value of $r_b$.
\section{Real Clusters}  \label{sec:Clusters}

In this section we demonstrate the techniques described in \S\ref{sec:infReg} 
by fitting SIM envelopes to a sample of three real clusters that have a tilted infall-caustic-like structure in redshift space. Our aim is to show the advantage of $\mbox{S}_v$ over $\mbox{S}_p$, to investigate how well a tilted SIM envelope matches the observed structure, and to estimate the possible peculiar velocities responsible for the observed tilt and other distortions. We also aim to show that tilted structures are found at a range of distances, not just nearby, and have chosen our sample to illustrate that.

The procedure of the work is as follows. 
(i) Plot a fat slice of a cluster field in $\mbox{S}_v$ to explore the structure of the cluster and its surroundings. (We choose the thickness of the slice to be the diameter of the cluster.)  
(ii) Orient the slice to align with the tilt, by rotating the slice direction until the apparent infall artifact is most tilted. 
(iii) Plot a thin slice of the cluster field in $\mbox{S}_v$ to see a cross section and show the caustic edges most sharply.  (We note the optimal thickness is around 2/3 $\alpha_{\text{turn}}$.) 
(iv) Select the value of $\sigma_{\text{vir}}$ from literature. (In a future work we will determine this value independently from our analysis.) 
(v) Determine the value of $v_{\text{obs}}$ that fits well with the cluster. 
(vi) Apply NDM to the thin slice to determine the best fit $\alpha_{\text{turn}}$ and $v_{0y}$ (or $v_{\text{circ}}$).  
(vii) Use the 3D envelope to determine the cluster's members from all the galaxies in the field.  
(viii) Compare the distribution of galaxies in $\mbox{S}_p$ and $\mbox{S}_v$.

The data sample is collected from SDSS-DR8 for the objects classified as galaxies. Visual morphological classification for nearly million galaxies from SDSS, which was the aim of the Galaxy Zoo project, has been published recently (see Lintott et al. 2008; Lintott et al. 2011; www.sdss.org \& www.galaxyzoo.org). Unfortunately, there is a cutoff range in $\delta$ for SDSS-DR8 which excludes the Virgo cluster, so we obtained the data for this cluster from the CfA ZCAT catalog (Huchra 2000).

\subsection{Virgo  ($z \sim 0.003$)}

The dynamical status and distances to galaxies in the Virgo cluster ($\alpha \approx 186.53^\circ$, $\delta \approx 12.86^\circ$) have been extensively studied in the literature (see e.g. Girardi et al. 1996; Ebeling et al. 1998; Rines \& Diaferio 2006; Mei et al. 2007; Karachentsev \& Nasonova 2010). In this section we investigate the distortion of the cluster in redshift space and show although SIM does not describe its appearance well in projection space $\mbox{S}_p$ it does a better job in slice space $\mbox{S}_v$. 

\subsubsection{Virgo in Redshift Space}

\begin{figure*}
\vspace{-0.4cm}
\includegraphics[width=29cm] {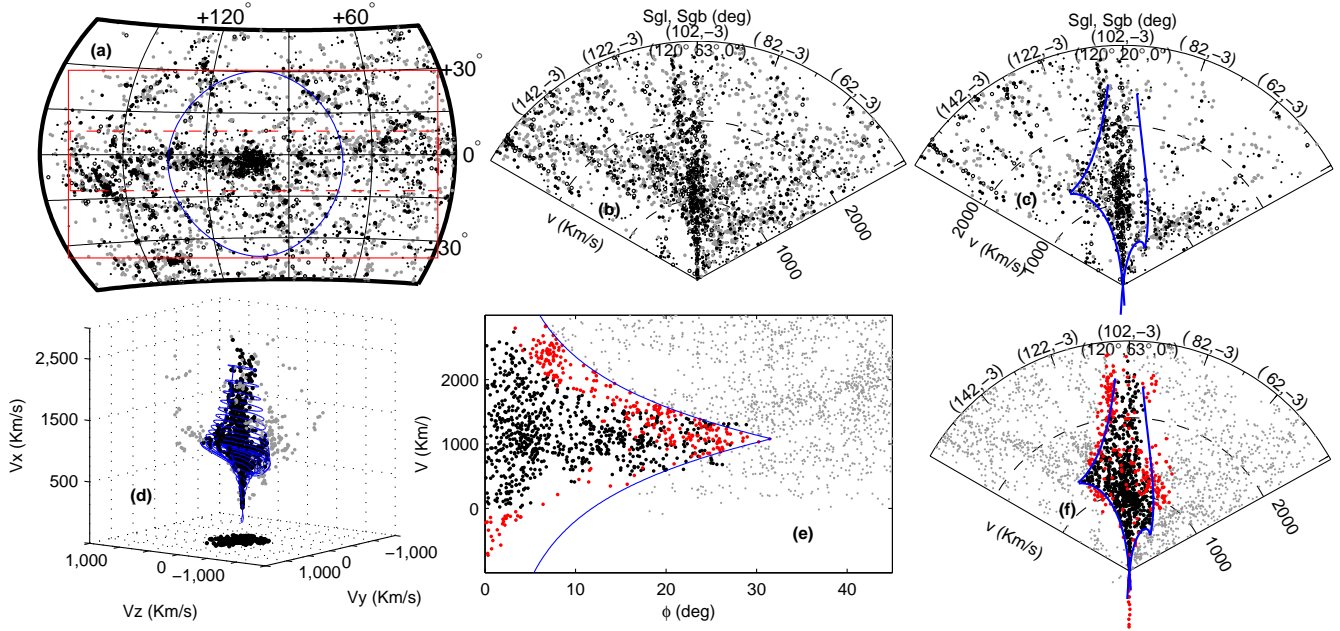} \vspace{-5cm}
\caption{Virgo. (a) Aitoff projection in supergalactic coordinates of elliptical ({\em big black}), dwarf elliptical ({\em open}), spiral ({\em gray}) and irregular ({\em small black}) galaxies with heliocentric velocities $< 3000 \mbox{ km s}^{-1}$.  
Circle is turnaround radius; rectangles outline $63^\circ$ thick fat slice ({\em solid line}) and $20^\circ$ thin slice ({\em dashed}).
(b) Fat slice in redshift space.  (c) Thin slice in redshift space, with PSM $\mbox{S}_v$ envelope.  (d) $\mbox{S}_v$ envelope in 3D. Members ({\em black}) are inside and outliers ({\em gray}) outside the envelope. (e) $\mbox{S}_p$ envelope, determined in projection space.  Members ({\em black}) and outliers ({\em gray}) from panel d are shown. False members ({\em red}) are outliers inside $\mbox{S}_p$ envelope.  (f) Fat slice showing positions of members, outliers, and false members from panel e.}
\label{fig:Virgo}
\end{figure*}

As usual for Virgo, galaxies are transformed from equatorial to supergalactic coordinates. Fig.~\ref{fig:Virgo}b shows $\mbox{S}_v$ of the fat slice to demonstrate the general features of the cluster and its surroundings. Fig.~\ref{fig:Virgo}c shows the distribution of the galaxies within the thin slice, in which we see the tilted caustic-shape pattern more sharply. As shown, the cluster's Finger appears clearly, since it includes most of the cluster's elliptical galaxies.  According to morphological segregation, the elliptical galaxies are settled in the virialized core of the clusters. This emphasizes that these galaxies, which appear to be extended in redshift space, are in fact fallen into the cluster's inner core and they just appear elongated because their large peculiar velocities distort their positions. 

Another distortion in the shape of the cluster is shown in panels b-c, where the left side of the cluster appears to go up while the right side goes down, and both sides are bounded by sharp curving edges. The nature of these edges (sharp and curving) suggests they are velocity caustics, which in turn suggests that the tilt could be due to transverse motion of the observer with respect to the cluster centre or flow with curl about the cluster centre (or both). 

We select $\sigma_{\text{vir}}=776\pm24$ km s$^{-1}$ (Rines \& Diaferio 2006) and determine $v_{\text{obs}}=1079$ km s$^{-1}$, which both fits the distribution of galaxies in the cluster field well and is used in the heliocentric frame in many references (e.g., Ebeling et al. 1998; Rines \& Diaferio 2006). 

The next step is to determine the cluster's turnaround angle $\alpha_{\text{turn}}$ and the observer's transverse relative motion $v_{0y}$ (or rotational flow of the turnaround $v_{\text{circ}}$) using NDM. Although there is another structure, the Ursa Major group, to the right of Virgo, its effect on the fitting of NDM is small (it makes little difference whether it is included or left out), so we apply NDM on the four sides of the cluster. The obtained parameters are $\alpha_{\text{turn}} =31.64\pm1.15^{\circ}$, $v_{0y}=-780\pm95$ km s$^{-1}$ (or $v_{\text{circ}}=-409\pm13$ km s$^{-1}$).

Using Eq.~\ref{eq:wtl}, the apparent width to length of the cluster is $\mathcal{W} = 0.42\pm0.02$. Since the intrinsic value is assumed to be $\mathcal{W}_0 = 0.46$ (PSM with $\Omega_0 = 0.27$), the infall towards the cluster is then $v_{0x}=94\pm24$ km s$^{-1}$ by Eq.~\ref{eq:v0x}, and the distance is $R =16.1\pm0.3$ Mpc by Eq.~\ref{eq:H0R} (with $h_0 = 0.73$).

For Hubble Space Telescope (HST) Key Project, Freedman et al. (2001), using Cepheids, find R=$14.6\pm0.3$ Mpc.   Mei et al. (2007) and Blakeslee et al. (2009), using the method of surface brightness fluctuations, find R = $16.5\pm1.2$ Mpc. Our estimate of R depends on $\sigma_{\text{vir}}$, $\alpha_{\text{turn}}$ and the choice of $\Omega_0$ and $h_0$. The value of $\sigma_{\text{vir}}$ that leads to R = 16.5 Mpc, fixing the other parameters, is 797 km s$^{-1}$. On the other hand the value of $\alpha_{\text{turn}}$ that gives the same R, fixing the other parameters, is $30.7^{\circ}$.

\begin{figure} 
\vspace{-0.5cm}
\includegraphics[width=21cm] {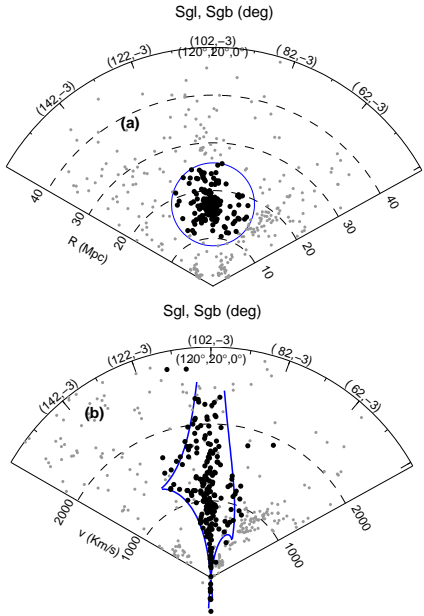} \vspace{-1.4cm}
\caption{The distribution of Virgo galaxies 
in (a) real space and (b) redshift space with SIM fit ({\em envelope}) and corresponding turnaround radius ({\em circle}) indicated. Black points are galaxies inside circle.
} 
\label{fig:Virgo1}
\end{figure}

After determining the cluster's parameters, a 3D envelope is used to obtain the cluster's members. As shown in Fig.~\ref{fig:Virgo}d, the cluster is clearly tilted in the 3D redshift space. The number of the galaxies ({\em black points}) located within the 3D envelope ({\em circled curves}) is 767 galaxies. Fig.~\ref{fig:Virgo}e shows the distribution of galaxies in $\mbox{S}_p$, where the black points indicate the cluster's members obtained from the 3D envelope and the two curves indicate the application of the PSM in $\mbox{S}_p$. Note that the caustic-shape pattern of the cluster is not clear in $\mbox{S}_p$ and the application of PSM can't describe the cluster in $\mbox{S}_p$. 
By contrast, a tilted caustic-shape pattern is very clear in $\mbox{S}_v$.  Although spherical infall such as PSM does not match this caustic-shape pattern well on the right side in $\mbox{S}_v$, it does describe the pattern well on the left side (see panel c; also see Fig. 3 in PS94).

The structure on the left side inside the envelope is the Virgo Southern Extension. The small finger outside the envelope centred at about 2500 km s$^{-1}$ immediately to the left of the main Virgo finger is the Virgo W cloud and is believed to be about twice as far away as the Virgo core (see, e.g., Mei et al. 2007).

The extended structure on the right side is the Ursa Major Group, centred at roughly $sgl \sim 62^\circ$ and 1000 km s$^{-1}$.  PSM does not describe this extended structure at all, of course, but even inside the region of the envelope the fit is not optimal. In particular, note that the top right edge of the {\em structure} is concave and seems to form part of a sweeping upward-facing curve over to Ursa Major but the top right side of the {\em envelope} is convex and curves downward, without extending as far as Ursa Major. The right top edge of the structure looks like a velocity caustic, but not one predicted by PSM.

\subsubsection{Virgo in Real Space} \label{sec:realvirgo}

The best way to explore the distortion of Virgo in redshift space is to plot the cluster in real space. Fortunately, the distances for 1797 galaxies within 3300 km s$^{-1}$ in the cluster field have been determined with accuracies ranging from about 10\% for ellipticals to 20\% for spirals (see Tully et al. 2008 (cosmicflows-1) \& Courtois et al. 2012).  Fig.~\ref{fig:Virgo1} shows the distribution of galaxies in real space (panel a) and redshift space (panel b).

The galaxies within the assumed turnaround radius in real space ({\em black points}) are explored in the redshift space. 
Notice these galaxies roughly fill the envelope.  This is different from the n-body cluster discussed earlier (Fig.~\ref{fig:simcls1}).
Specifically, the galaxies on the left side fit pretty well with the left edge of the envelope. Also note that the small finger on the upper left side in Fig.~\ref{fig:Virgo}c (the Virgo W cloud, mentioned above) does not appear in these plots, because its more distant galaxies are not in the catalog. This both confirms that the small finger does not belong to the cluster and also suggests that a SIM such as PSM can describe the infall pattern of Virgo on the left side.

On the right side, however, a number of galaxies fall outside the envelope on the right top edge and the outline of the overall structure looks different from the envelope. As we discussed above, this indicates that a simple SIM such as PSM can't describe the infall pattern of Virgo on right side. However, since the right top edge looks like a velocity caustic, it seems likely some other type of flow is producing this structure.

Also note that this group of galaxies, which our analysis in Fig.~\ref{fig:Virgo} (panels e and f) would classify as `false members' of the cluster since they fall inside the $\mbox{S}_p$ envelope but outside the 3D $\mbox{S}_v$ envelope, are actually true members, since they fall within Virgo's turnaround radius in real space.  However, other false members identified in Fig.~\ref{fig:Virgo} are truly outside the turnaround, the most obvious being the W cloud. This illustrates the problem with using $\mbox{S}_p$ to identify members. In a future work we will study the infall pattern of Virgo and compare the distribution of galaxies in real and redshift spaces in detail.

\subsection{A1459 ($z \sim 0.02$)}

\begin{figure*} \vspace{-0.3cm}
\includegraphics[width=19cm] {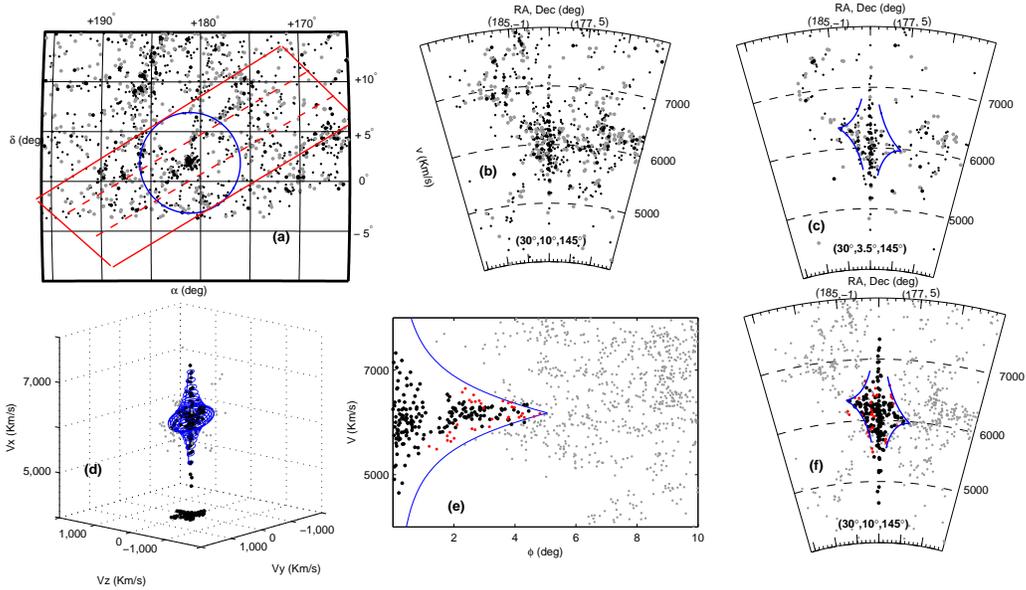} \vspace{-1.2cm}
\caption{A1459. Similar to Fig.~\ref{fig:Virgo}.  (a), (b), \& (c) Aitoff projection in celestial coordinates,  $10^\circ$ thick fat slice, and $2^\circ$ thin slice, respectively, of elliptical ({\em big black}), spiral ({\em gray}) and irregular ({\em small black}) galaxies.  (d) $\mbox{S}_v$ envelope in 3D, with members ({\em black}) 
and outliers ({\em gray}). (e) \& (f) Members and outliers from panel d, in $\mbox{S}_p$ and fat slice, respectively.  False members ({\em red}) are outliers inside $\mbox{S}_p$ envelope.}
\label{fig:A1459}
\end{figure*}

Aguerri, Sa\'nchez-Janssen \& M\"unoz-T\"un\'on (2007) and AA11 have showen that A1459 ($\alpha = 180.43^{\circ}$, $\delta = 2.78^{\circ}$) is located at $\approx 6100$ km s$^{-1}$. Fig.~\ref{fig:A1459} is similar to Fig.~\ref{fig:Virgo} in which the distribution of galaxies is shown in the field of A1459 in $\mbox{S}_v$ and $\mbox{S}_p$. 
The orientation that shows the tilted caustic-shape pattern clearly is $145^{\circ}$. Note that the cluster's Finger, which contains most of the elliptical galaxies in the cluster, shows up clearly. Also note that shape of the cluster is sharper in the thin slice and the bounding edges are caustic-like.

The velocity dispersion of the cluster at the virial radius $r_{\text{vir}}$ is $\sigma_{\text{vir}}=516\pm56$ km s$^{-1}$ (AA11). We select $v_{\text{obs}}=6180$ km s$^{-1}$ that is fitted well with the distribution of galaxies in the field of the cluster. The application of NDM gives the best fit $\alpha_{\text{turn}} = 5.1\pm0.3$ and $v_{0y} = 2260\pm341$ km s$^{-1}$ (or $v_{\text{circ}} = 199\pm13$ km s$^{-1}$).

Using Eq.~\ref{eq:wtl}, the apparent width to length ratio of the cluster is $\mathcal{W}=0.61\pm0.05$. If the intrinsic value is $\mathcal{W}_0 = 0.46$ (PSM with $\Omega_0 = 0.27$), by Eqs.~\ref{eq:v0x} and \ref{eq:H0R} (with $h_0 = 0.73)$ our radial velocity with respect to the cluster is $v_{0x}=-1542\pm210$ km s$^{-1}$ (so, away from the cluster) and the cluster distance is R $=63.5\pm2.9$ Mpc .

Fig.~\ref{fig:A1459}d shows the application of the 3D envelope to determine the cluster members ({\em black points}). Fig.~\ref{fig:A1459}e shows the distribution of the 173 cluster's members ({\em black points}) that determined from the 3D envelope in panel (d) and outliers ({\em gray points}) that fall out of the 3D envelope. The two curves shown in panel (e) refer to the application of PSM in $\mbox{S}_p$. 

Note that not all the galaxies at the outer region of the cluster are members. They appear to make a substructure in this region but this is due to the projection effect, which convolves the galaxies around the axis, putting the galaxies on different sides of the cluster on top of each other and obscuring any tilt, such as that due to the transverse peculiar velocity of the observer and/or rotational motion about the cluster centre. 

Because the structure of A1459 is in fact tilted, a SIM such as PSM can't describe the cluster well in $\mbox{S}_p$ (Fig.~\ref{fig:A1459}d). By contrast, PSM can describe the infall pattern of the cluster pretty well in $\mbox{S}_v$ (Fig.~\ref{fig:A1459}c).

Fig.~\ref{fig:A1459}e and f show the positions in $\mbox{S}_p$ and $\mbox{S}_v$ of the false members ({\em red points}). Again we note that using $\mbox{S}_p$ to identify cluster members leads to an overcount, compared to the members identified using the 3D $\mbox{S}_v$ envelope.
\subsection{A1066 ($z \sim 0.07$)}

\begin{figure*} \vspace{-0.3cm}
\includegraphics[width=18cm] {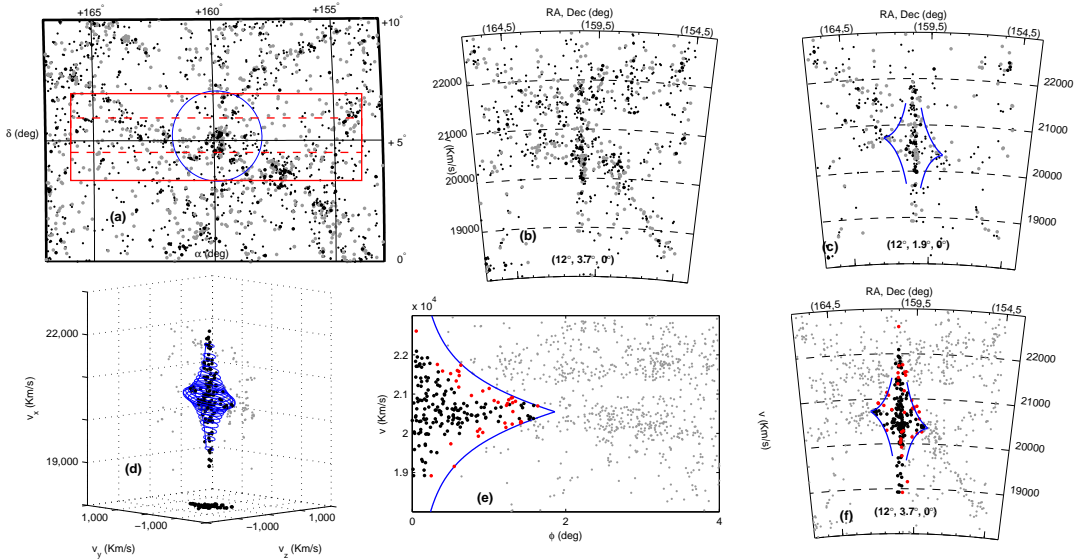} \vspace{-1.2cm}
\caption{A1066.  Same as Fig.~\ref{fig:A1459}.  The fat and thin slices (panels b and c) have thicknesses $3.7^\circ$ and $1.5^\circ$.} 
\label{fig:A1066}
\end{figure*}

The cluster A1066 ($\alpha = 159.88^\circ$, $\delta = 5.18^\circ$; AA11) has been studied in literature in $\mbox{S}_p$ (see e.g., Rines \& Diaferio 2006; Aguerri, Sa\'nchez-Janssen \& M\"unoz-T\"un\'on 2007; Yoon et al. 2008; AA11). Fig.~\ref{fig:A1066} is the same as Figs.~\ref{fig:Virgo} and \ref{fig:A1459} and shows the difference between $\mbox{S}_v$ and  $\mbox{S}_p$. 
The finger is shown clearly in Figs.~\ref{fig:A1066}b \& ~\ref{fig:A1066}c from the distribution of elliptical galaxies. A tilted caustic-shaped pattern shows up clearly in the thin slice ($1.5^{\circ}$ thick).

The velocity dispersion of the cluster at $r_{\text{vir}}$ is $\sigma_{\text{vir}}=764\pm80$ km s$^{-1}$ (AA11). The value $v_{\text{obs}}=20550$ km s$^{-1}$ gives a good fit with the distribution of galaxies in the cluster field and is in the range of the radial velocities of the cluster posted in literature (see e.g., Rines \& Diaferio 2006; Aguerri et al. 2007; Yoon et al. 2008; AA11). Applying NDM to the thin slice gives $\alpha_{\text{turn}}=1.86\pm 0.14^{\circ}$ and $v_{0y}=-6130\pm254$ km s$^{-1}$ (or $v_{\text{circ}}= -199\pm15$ km s$^{-1}$). 

The apparent width to length of the cluster is $\mathcal{W}=0.50\pm0.04$. If the intrinsic value is $\mathcal{W}_0 = 0.46$ as we've assumed for the other clusters, then our radial velocity with respect to the cluster is  $v_{0x}=-1889\pm721$ km s$^{-1}$ (away from the cluster) and the cluster distance is R$=255.6\pm9.9$ Mpc.

Some of the cluster's members (156 galaxies) which are determined from the 3D envelope (panel d) are outside the boundary of PSM in $\mbox{S}_p$ (panel e). On the other side some of the cluster's members obtained from the application of PSM in $\mbox{S}_p$ (panel e) are outside of the envelope of the cluster (panel f). The reason is again due to putting the right and left sides of the clusters on each other, the projection effect in $\mbox{S}_p$ and disappearing of the effect of the transverse peculiar velocity of the observer in its rest frame or rotational flow of the cluster's shells or both in $\mbox{S}_p$.

\section{Discussion of Cluster Fits}
\label{sec:Disc}

As shown  in \S\ref{sec:Clusters}, Virgo, A1459, and A1066 are examples of galaxy clusters with redshift space structures that are both bounded by curved concave edges on four sides and {\em tilted}.  Consequently, it is possible to fit all three clusters with spherical infall, provided one allows the observer to have transverse motion relative to the cluster.

The SIM we chose to use was PSM, for convenience, but that choice has little effect on the findings. It is also not necessarily the case that transverse motion causes the observed tilt.  However, using it is an easy way to generate a tilted envelope for the fit, which can then be used to measure the amount of tilt, which in turn allows an estimate of the magnitude of other flow, such as circular velocity about the cluster centre, that could produce that amount of tilt.

Tables \ref{tab:independent} and \ref{tab:dependent} sum up the parameters of the three clusters Virgo, A1459, and A1066 analyzed this way.  Table~\ref{tab:independent} contains the model independent parameters, namely, `finger length'  measured two ways (via virial velocity dispersion or crude direct measurement; columns 2--5);
cluster velocity, angular size of turnaround, and slope of tilt (columns 6--9); and transverse velocity or circular velocity implied by the tilt (columns 10--11).  Table~\ref{tab:dependent} contains the observed width to length ratio (which depend on finger length measurement choice; columns 2 \& 7) and the 
two model dependent parameters, radial velocity towards the cluster and distance (columns 3--6 \& 8--11). The parameters are discussed in detail below.

\begin{table*} \centering
\caption{Model-{\em independent} parameters of each cluster. The parameters in bold determine $v_{0y}$ and $v_{circ}$.}
\label{tab:independent}
\scriptsize
\begin{tabular}{ccccccccccc}\hline
&$\sigma_{\text{vir}}(ref)$&$L$&$L/\sigma_{(ref)}$&$\sigma(L)$&$v_{\text{obs}} (ref)$&$\mathbf{v_{obs}(fit)}$&$\mathbf{\alpha_{turn}}$&{\bf slope}&$v_{0y}$&$v_{\text{circ}}$\\
&(km s$^{-1}$)&(km s$^{-1}$)&&(km s$^{-1}$)&(km s$^{-1}$)&(km s$^{-1}$)&($^\circ$)&&(km s$^{-1}$)&(km s$^{-1}$)\\
\hline
Virgo&$776\pm24^a$&3313&4.3&869&$1079^{a,b}, 1035^c$&{\bf 1079} &{\bf 31.64$\pm$1.15}&{\bf -0.72$\pm$0.09}&-780 $\pm$95 &-409$\pm$13\\
A1459&$516\pm56^d$&2675&5.2&702&$6090^d, 6010^e$&{\bf 6180} &{\bf 5.06 $\pm$0.33}&{\bf 0.37$\pm$0.06}&2260 $\pm$341 &199 $\pm$13\\
A1066&$764\pm80^d$&3170&4.1&832&$20506^a, 20657^d, 20708^e$&{\bf 20550}&{\bf 1.86 $\pm$0.14}&{\bf -0.30$\pm$0.01}&-6130$\pm$254&-199$\pm$15\\
\hline
\end{tabular}

\footnotesize{a = Rines \& Diaferio (2006), b = Ebeling et al. (1998), c = Mould et al. (2000), d = Abdullah et al. (2011), e = Aguerri, Sa\'nchez-Janssen \& M\"unoz-T\"un\'on (2007)}
\end{table*}

\begin{table*} \centering
	
\caption{Model-{\em dependent} parameters of each cluster.}
\label{tab:dependent}
\scriptsize
\begin{tabular}{ccccccc cccc ccc}\hline
&\multicolumn{6}{c}{$\sigma = \sigma_{\text{vir}}(ref)$}&&\multicolumn{5}{c}{$\sigma = \sigma(L)$}\\
\cline{2-7} \cline{9-14}\\ [-5pt]
&&\multicolumn{2}{c}{$\mathcal{W}_0 = \mathcal{W}_{PSM}$}&&\multicolumn{2}{c}{$\mathcal{W}_0 = \mathcal{W}_{1066}$}&&&\multicolumn{2}{c}{$\mathcal{W}_0 = \mathcal{W}_{PSM}$}&&\multicolumn{2}{c}{$\mathcal{W}_0 = \mathcal{W}_{1066}$}
\\
\cline{3-4} \cline{6-7} \cline{10-11} \cline{13-14}\\ [-5pt]           
&$\mathcal{W}$&$v_{0x}$& $R$&&$v_{0x}$&$R$&&$\mathcal{W}$&$v_{0x}$&$R$&& $v_{0x}$&$R$      \\
&&(km s$^{-1}$)&(Mpc)&&(km s$^{-1}$)& (Mpc) &&&(km s$^{-1}$)&(Mpc)&&(km s$^{-1}$)&(Mpc)\\
\hline                                                       
Virgo&0.42$\pm$0.02&94   $\pm$53  &16.1 $\pm$0.7 &&$213  \pm58$ &$17.7\pm 0.8$ 
&&0.38$\pm0.01$ & 235$\pm$43 & 18.0$\pm$0.8 && 250$\pm$43 & 18.2$\pm$0.6
\\
A1459&0.61$\pm$0.08&-1542$\pm$587&63.5 $\pm$8.0 &&$-1072\pm646$&$70.0\pm9.0$
&&0.45$\pm$0.03 & 130$\pm$411 & 86.4$\pm$5.6 && 201$\pm$415 & 87.4$\pm$5.7
\\
A1066&0.50$\pm$0.07&-1889$\pm$2406&255.6$\pm$33.0&&0&$281.5\pm0$
&&0.46$\pm$0.04 & -230$\pm$1529 & 278.4$\pm$20.9 && 0 & 281.5$\pm$0
\\
\hline
\end{tabular}
\end{table*}

\subsection{\textbf{Velocity dispersion $\sigma$ \& finger length $L$}}
\label{subsec:sigma}

Columns 2--5 of Table~\ref{tab:independent} give parameters related to finger length of the clusters. 
Column~2  gives the virial velocity dispersion $\sigma_{vir}(ref)$, as measured in a reference; column~3 gives the length $L$ of the finger of each cluster, as measured by us; and Column 4 gives the ratio $L/\sigma_{vir}$.  Column 5 gives a dispersion proportional to $L$, explained below, defined $\sigma(L) \equiv L/(2.2\sqrt{3})$. 

The main thing to note here is that the ratio $L/\sigma_{vir}$ is not the same from cluster to cluster.  It is larger for one cluster (A1459) than for the other two by about $\sim24\%$. So, is it better to define the length of the artifact using virial velocity dispersion $\sigma_{vir}$ (as we did in fitting the envelope) or using the directly measured length $L$?

To answer this question, it is helpful to understand the reason for the difference between $\sigma_{vir}$ and $L$, and also recall why we are interested in the length of the finger.  

In the literature, the velocity dispersion at the virial radius $\sigma_{vir}$ is defined as the standard deviation about the mean of the observed velocities of all galaxies within that projected radius  (see e.g., Fadda et al. 1996; Girardi et al. 1996; AA11).  We measured finger length $L$ by plotting very thin core-centred slices of 4.0, 0.3 and 0.2 $^\circ$ for Virgo, A1459 and A1066, respectively and taking the difference between the smallest and largest velocity in the finger. Looking at $\text{S}_p$ in Figs.~\ref{fig:Virgo}e, \ref{fig:A1459}e, and \ref{fig:A1066}e, we can see that the velocities of core galaxies in A1459 are more tightly clustered about the mean observed velocity (giving it a smaller $\sigma_{vir}$, proportionately), perhaps because A1459 is more virialized than Virgo or A1066.

Studies have shown that well-virialized cluster cores show a morphological separation in velocity dispersion, with  late type spirals having markedly larger velocity dispersion than ellipticals or early type spirals, probably because they have only recently fallen in and their orbits are still radial, with the velocity of infall, while the others have virialized 
(see, e.g.,  Adami, Biviano \& Mazure 1998).   
This could be the situation in A1459, that a smaller percentage of galaxies have the velocity of infall compared to Virgo and A1066.  
Note that in each case, it is these unvirialized galaxies that determine $L$.

Now, what do we need finger length for?  As mentioned above, we set the `$\sigma$' in PSM to  $\sigma = \sigma_{vir}(ref)$ in fitting an envelope.  However, we find that choice has essentially no effect on other parameters in Table~\ref{tab:independent}.   Boosting $\sigma$ up or down in the fit envelope does not change the observed velocity,  angular size of the turnaround, or slope of the fit, and thus does not change the estimated transverse velocity $v_{0y}$ or circular flow $v_{circ}$ associated with those fit parameters.

This does not mean that finger length is unimportant.  We need it to calculate the observed width to length ratio $\mathcal{W}$, which in Eq.~\ref{eq:wtl} is defined in terms of $\sigma_{vir}$.  The two model-dependent parameters, radial velocity towards the cluster $v_{0x}$ and cluster distance $R$, depend directly on $\mathcal{W}$, as well as on the assumed intrinsic ratio $\mathcal{W}_0$.

It seems possible that the correct $\sigma_{vir}$ to use in calculating $\mathcal{W}$ is the dispersion of late type spirals only, since that reflects the strength of the infall field near the core.  Lacking that, we instead define a `dispersion' $\sigma(L)$ proportional to the measured finger length $L$,  which may in turn be proportional to the dispersion of spirals. That way we can compare with results that use $\sigma_{vir}(ref)$.

The definition we choose is $\sigma(L)  \equiv L/(2.2\sqrt{3})$. For comparison, a finger of length $L$ with a completely uniform distribution of 
 observed velocities has a velocity dispersion of $\sigma = L/(2 \sqrt{3})$, as can be verified by, e.g., numerical simulation.  We chose this particular definition (with a factor of 2.2 rather than 2 or some other value) for reasons we will discuss below, in \S\ref{subsec:v0xR}.

\subsection{\textbf{Observed velocity} $v_{\text{obs}}$}

Columns~6 and 7 in Table~\ref{tab:independent} give $v_{obs}(ref)$, the cluster radial velocity selected from references, and $v_{obs}(fit)$, the velocity of the best fit envelope.

We treated $v_{obs}(fit)$ as a subjectively determined input in the fit, because the current form of NDM does not work well finding the best $v_{obs}$.  Therefore, we determined this by eye,  and this is the reason there is no associated standard deviation uncertainty.

We find that the radial velocity of the best fit envelope of Virgo is the same as $v_{\text{obs}} (ref)$. This value, $v_{\text{obs} (fit)} \approx 1079$ km s$^{-1}$ is widely used in the literature (see e.g. Sandage \& Tammann 1976) with respect to heliocentric rest frame. However, $v_{\text{obs}} (fit)$ of A1459 and A1066 are different from that posted in the literature. An interpretation of the difference is that $v_{obs}(ref)$ for these two clusters was found by finding mean velocity in $\mbox{S}_p$ without taking the problems of this redshift space into consideration.

\subsection{Transverse velocity $v_{0y}$ \& circular velocity $v_{\text{circ}}$}

Columns~8 \& 9 of Table ~\ref{tab:independent} give the angular size and slope of the best fit envelope, and Columns~10 \& 11 give the transverse motion of the observer relative to the cluster $v_{0y}$ or circular velocity at the turnaround radius $v_{circ}$ that produce that amount of tilt, from  Eq.~\ref{eq:v0yvcirc}.

Recall Virgo, A1066, and A1066 lie at redshifts $z \sim $ 0.003, 0.02, and 0.07, respectively.  If the tilts are produced only by relative transverse velocity between observer and cluster we expect tilt should in general decrease with increasing redshift (see Eq.~\ref{eq:v0yvcirc}), assuming transverse velocities of reasonable size ($< \sim1000$ km s$^{-1}$). The magnitudes of the slopes (0.72, 0.37, 0.30) do decrease with redshift, but the magnitudes of the transverse velocity required to produce those slopes (780, 2260, 6130 km s$^{-1}$) grow rapidly with distance to unreasonable size.  

Conversely, if we assume $v_{0y} = 0$ and estimate the circular flow needed to produce the observed tilt, we find the magnitudes of these values (409, 199, 199 km s$^{-1}$) to be reasonable.  

One can also investigate a combination of the two motions, $v_{\text{circ}}$ and $v_{0y}$, where the cluster may have a rotational flow in additional to its transverse motion with respect to the observer. Suppose the three clusters experience the same magnitude of rotational flow $v_{\text{circ}} = 185$ km s$^{-1}$. Using the slopes listed in Table~\ref{tab:independent} in Eq.~\ref{eq:slope},  the magnitudes of the transverse velocities required are  $v_{0y} = 427, 163$ and $430$ km s$^{-1}$ for Virgo, A1459 and A1066, respectively (reasonable values). 

What about other possible causes of the tilt?  For example, could the tilt just be due to real structure, such as a filament leading into and out of the cluster that is inclined to the line of sight?

Such a scenario is certainly possible.  
The n-body cluster discussed earlier (Fig.~\ref{fig:simcls1}) has a structure in redshift space that looks similar to the tilted structures seen in survey data, and in that case the tilt is due to real structure, not velocity distortion.  We do not have distance information for galaxies in our two distant clusters (A1459 and A1066) so this possibility cannot be ruled out for them and may in fact be what is going on (compare Figs.~\ref{fig:A1459}c and \ref{fig:A1066}c to Fig. \ref{fig:simcls}d).  

\begin{figure*}
\vspace{-0.3cm}
   \centering
\hspace*{-1cm}
   \includegraphics[width=37cm]{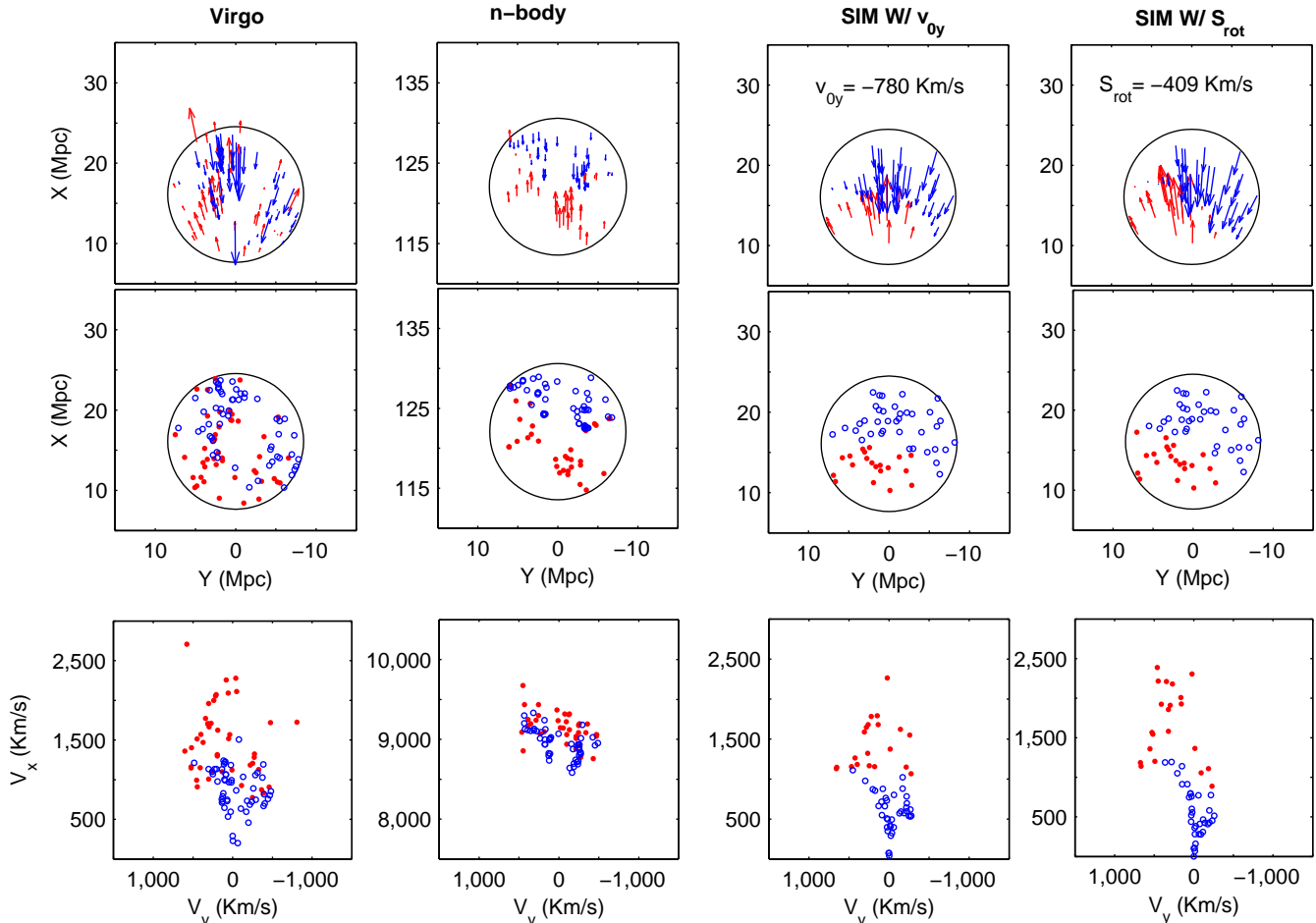} \vspace{-5cm}
     \caption{Galaxies with $r_{vir} < r < r_{turn}$ for Virgo, 
  `tilted' n-body cluster (Fig.~\ref{fig:simcls1}), and toy spherical infall models with either transverse or rotational motion. Turnaround is that assumed for SIM envelope fit.  ({\em Top}) Peculiar velocities towards ({\em blue}) or away ({\em red}) from us.   ({\em Middle \& bottom}) Real space and redshift space positions of galaxies with velocity towards us ({\em blue open circles}) or away ({\em red filled points}).}
   \label{fig:VirNbodSim}
\end{figure*}

However, this is not what is going on in the Virgo cluster.  Fig.~\ref{fig:VirNbodSim} show the peculiar velocities of those Virgo galaxies that have measured distances, in the infall region $r_{vir} < r < r_{turn}$, where the turnaround radius is that assumed for the SIM envelope fit.  The `tilted' n-body cluster is also shown for comparison, as well as two toy spherical infall models, one tilted by transverse velocity $v_{0y}  = -780$ km s$^{-1}$ and the other by constant rotational speed $s_{rot} = -409$~km~s$^{-1}$.

The radial components of Virgo's peculiar velocities towards or away from us are different from the n-body cluster.  They are larger in magnitude and also show a pattern that the n-body cluster's do not, going predominantly away from us on the left side and towards us on the right side.
This is similar to the pattern seen in the two toy SIMs with transverse motion or rotational flow.  This is also similar to the recent PANDAs result finding dwarf galaxies on one side of Andromeda Galaxy are coming towards us while those on the other side are going away, suggestive of rotational flow (Ibata et al. 2013).

\subsection{\textbf{Radial velocity} $v_{0x}$~\&~distance~$R$}

\label{subsec:v0xR}

In PSM, every infall region, large or small, shares the same intrinsic width-to-length ratio $\mathcal{W}_0$. This may not be true for real clusters, but if it is, one can find the deviation of the observed length-to-width ratio $\mathcal{W}$ from $\mathcal{W}_0$, and then determine the radial peculiar velocity of the observer with respect to a cluster centre, $v_{0x}$, and the distance to the cluster, $R$.  Table~\ref{tab:dependent} summarizes the dependence of $\mathcal{W}$, $v_{0x}$ and $R$, on choice of velocity dispersion $\sigma$ and assumed $\mathcal{W}_0$.

\subsubsection*{ $\bullet$ $\sigma = \sigma_{vir}(ref)$}

In the first half of Table \ref{tab:dependent} (columns 2--6) we calculate $\mathcal{W}$, $v_{0x}$ and $R$ from the parameters of the best fit envelope listed in Table \ref{tab:independent} and $\sigma_{\text{vir}} (ref)$ taken from the literature.

We use  Eq.~\ref{eq:wtl} to find $\mathcal{W}$ (column 2), then assume $\mathcal{W}_0 = \mathcal{W}_{PSM} = 0.458$ and find $v_{0x}$ and $R$ (columns 3--4) using 
Eqs.~\ref{eq:v0x} and \ref{eq:H0R}. Also, as discussed in \S\ref{sec:wtl} and using Eqs.~\ref{eq:v0x} and \ref{eq:H0R}, we determine $v_{0x}$ and $R$ for Virgo and A1459, by assuming $\mathcal{W}_0 = \mathcal{W}_{A1066} = 0.50$, where $\mathcal{W}_{A1066}$ is the width-to-length ratio of our most distant cluster (columns 5--6).

In the first case ($\mathcal{W}_0 = \mathcal{W}_{PSM}$), one can notice that $v_{0x}$ (column 3) is large (not reasonable) for the two clusters A1459 and A1066. Here we depend on $\sigma_{\text{vir}} (ref)$ that was determined from the traditional different methods for membership selection applied in $\mbox{S}_p$. This may affect the value of $\sigma_{\text{vir}} (ref)$ and then certainly affects the value of $v_{0x}$. However, when we assume $\mathcal{W}_0 = \mathcal{W}_{A1066}$, as in the second case, $v_{0x}$ becomes smaller. 

Another parameter that $v_{0x}$ depends on is $\alpha_{\text{turn}}$. Its effect is very sensitive for distant clusters. I.e., as the distance to the cluster increases, a small change in the angular separation causes a big change in $v_{0x}$. For A1066, for example, $\alpha_{\text{turn}} = 1.86 ^{\circ}$, this gives $v_{0x} = - 1886$ km s$^{-1}$. A small decrease in $\alpha_{\text{turn}}$ by $0.1^{\circ}$ gives $v_{0x} = - 829$ km s$^{-1}$. 

Finally, as $v_{0x}$ depends on $\alpha_{\text{turn}}$ and $\sigma_{\text{vir}}$, we can reduce the value of $v_{0x}$ for A1459 and A1066 by assuming a little bit change in both $\alpha_{\text{turn}}$ and $\sigma_{\text{vir}}$. For A1459, assuming $\alpha_{\text{turn}} = 4.7^{\circ}$ and $\sigma_{\text{vir}} = 580$ km s$^{-1}$, $v_{0x} = -568$ km s$^{-1}$, while for A1066, $v_{0x} = -580$ km s$^{-1}$ for $\alpha_{\text{turn}} = 1.82^{\circ}$ and $\sigma_{\text{vir}} = 800$ km s$^{-1}$. Note that a little bit change in $\alpha_{\text{turn}}$ and $\sigma_{\text{vir}}$ causes a radical change in $v_{0x}$. This shows that $v_{0x}$ is very sensitive to the other parameters which should be determined carefully. 

\subsubsection*{ $\bullet$ $\sigma = \sigma(L)$}

In the second half of Table~\ref{tab:dependent} we calculate $\mathcal{W}$, $v_{0x}$, and $R$ (columns 7--11) as before, but with $\sigma(L)$ from Table~\ref{tab:independent}, where $\sigma(L) \equiv L/(2.2 \sqrt{3})$ is proportional to finger length (see \S\ref{subsec:sigma}).

The values of $\mathcal{W}$ (column 7) are 0.38 (0.3761), 0.45 (0.4483), and 0.46 (0.4629) for Virgo, A1459, and A1066, respectively. Here are some things to note about these values.  (1)~The reason the definition of $\sigma(L)$ has a factor of 2.2, rather than 2 as for a uniform distribution (\S\ref{subsec:sigma}), is to  make $\mathcal{W}$ for A1066 (0.46) close to the value of $\mathcal{W}_{\text{PSM}}$ (0.458). (2)~The values  show less variation as redshift increases, which is what we expect (see Eq.~\ref{eq:deltaWW0}), and which is a trend independent of the proportionality constant in $\sigma(L)$.  (3)~The values of $\mathcal{W}$ using $\sigma_{vir}(ref)$  (0.42, 0.61, and 0.50) do not show this trend.

The values of $v_{0x}$ and $R$ when $\mathcal{W}_0 = \mathcal{W}_{PSM}$ and $\mathcal{W}_0 = \mathcal{W}_{1066} = 0.4629$ are given in columns (8--9) and (10--11), respectively. These are calculated using  unrounded values of $\mathcal{W}$.

There are a couple things to notice here. First, the values of $v_{0x}$ in each case are small (reasonable):  when $\mathcal{W}_0 = \mathcal{W}_{PSM}$  the values are 235, 130, and -230 km s$^{-1}$ for Virgo, A1459, and A1066, respectively (column 8); and when $\mathcal{W}_0 = \mathcal{W}_{1066}$, the values are 250, 201, and 0 km s$^{-1}$.

Second, notice that the values of $v_{0x}$ for Virgo and A1459 are similar in the second case. In fact, if the assumed value of $\mathcal{W}_0$ is increased to be slightly larger than $W_{1066}$, the values of $v_{0x}$ will be similar for all three clusters. For example, if $\mathcal{W}_0 = 0.467$, the velocities are 261, 258, and 182 km s$^{-1}$. This is because now (with this assumed $\mathcal{W}_0$) all three clusters have an apparent width which is less than it `should' be, as caused by an observer moving toward each one of them (see bottom panels of Fig.~\ref{fig:posnegV0x}).

This is interesting because Virgo, A1459, and A1066 all lie in approximately the same direction.  Virgo and A1459 are separated by $\sim 12^\circ$, A1459 and A1066 by $\sim20^\circ$, and A1066 and Virgo by $\sim28^\circ$.  So, if the peculiar velocity of the observer is much larger than that of any of the cluster centres, as might be the case, and if they lie in a similar direction, then we expect that $v_{0x}$ (the component of the observer's peculiar velocity with respect to a cluster centre that points toward that cluster)  will not differ much from cluster to cluster.

Note we are working in the heliocentric frame and our motion with respect to the CMB in this frame is 371.9 km s$^{-1}$ towards $\alpha = 168.01^\circ$, $\delta = -6.98^\circ$ (Fixsen et al. 1996). This direction is $\sim27^\circ$ from Virgo, $\sim16^\circ$ from A1459, and $\sim15^\circ$ from A1066, and the component of our heliocentric CMB motion towards each of these clusters is approximately $331$, $358$, and $359$ km s$^{-1}$ respectively.  This is not greatly different from values calculated above, and if we increase the value of $\mathcal{W}_0$ still more we can cause the calculated value of $v_{0x}$ for A1066 to exceed the other two, since an increase in $\mathcal{W}_0$ causes the most distant cluster's $v_{0x}$ to increase most rapidly. For example, if $\mathcal{W}_0 = 0.470$, the velocities are 269, 299, and 315 km s$^{-1}$.  

However, as noted before, the uncertainties in these calculations are large, becoming very large as distance increases, and we only have three clusters in this demonstration.  So, we would not say this proves anything.  However, it does seem to suggest that using the velocity dispersion of late type spirals (or lacking that, a `dispersion' proportional to finger length such as $\sigma(L)$), and not the virial dispersion $\sigma_{vir}$, to calculate the width-to-length ratio $\mathcal{W}$ is the correct thing to do.

\section{Summary}

\label{sec:Con}

We have shown there are three clusters  (Virgo, A1459, and A1066) whose shapes in redshift space resemble tilted infall artifacts.  We've shown these shapes can be fitted by spherical infall models (SIMs) that include transverse motion between observer and cluster or shear flow such as rotation. 
Because of the tilt, the characteristic two-trumpet-horn shape is apparent in slice space $\mbox{S}_v$, but obscured in axially convolved projection space $\mbox{S}_p$.  Since most studies of clusters have been in $S_p$ (see  \S\ref{sec:Intro}), such tilt may be more common than currently recognized.
 
Past studies of cosmological 
 simulations have found that SIMs do not describe n-body clusters in $\mbox{S}_p$ (\S\ref{sec:Intro}), and our explorations find this seems to be true in $\mbox{S}_v$ as well.   Even in slices, we saw little evidence of infall distortion, tilted or otherwise, around clusters in simulations. However, we did find one n-body cluster that looks, in redshift space, like the tilted clusters seen in survey data (\S\ref{sec:infReg}).  This structure is not an infall artifact but mostly real (a pseudo-artifact).

As we've discussed, we don't have peculiar velocities for the more distant two of our three clusters  and so cannot easily tell if they are actual infall artifacts or pseudo-artifacts like the one example we found in simulation.  However, the velocity field of the nearest tilted cluster (Virgo) differs considerably from the pseudo-artifact.  Instead, it resembles the SIM toy models, as we've shown. 

This suggests that SIMs should not be ruled out just because they do not match current n-body simulations.  They could be a useful tool for both rough analysis of survey data and for testing simulation (as we've done here). The following summarizes the techniques developed in this paper  for using SIMs in this way.

\paragraph*{1.} Searches for infall distortion in survey data or simulation should be done in $\mbox{S}_v$, not $\mbox{S}_p$.  As we have shown, assuming axial symmetry leads to problems when tilt is present.
 
Plotting galaxy clusters in $\mbox{S}_v$ also avoids other defects of $\mbox{S}_p$ detailed in  \S\ref{sec:SpSv}, since plots in $\mbox{S}_v$ show the true features of galaxy clusters and do not depend on choice of clusters' centres.

\paragraph*{2.} The SIM-based 2D and 3D envelopes introduced here to describe the tilted caustic-shape pattern of galaxy clusters in redshift space can be exploited to obtain a cluster's members, which is one of the important keys in studying dynamics of galaxy clusters. 

The 3D envelope we use is the first attempt that uses 
three dimensions to avoid the projection effect which is the main problem of all previous methods used to constrain clusters members. We apply the 3D envelope on toy spherical infall simulated clusters with known true members to test this method and obtained good results. The only factor that affects this method is the triple-value problem (see Tonry \& Davis 1981). 

\paragraph*{3.} The new algorithm (NDM) introduced here obtains the best envelope that matches with the distribution of galaxies in a cluster field by determining number density within the envelope. Previously, the choice of the best envelope was performed manually, depending on eye, which may cause different results for the same cluster.  Our algorithm is not fully automated---there are still some parameters that must be set manually.  However, application of NDM tested on 10 simulated clusters gives promising results.

\paragraph*{4.}  The analysis techniques introduced here utilizing the tilt and the width to length ratio $\mathcal{W}$ of the infall artifact (as determined by a fitted SIM envelope) can potentially be used to study peculiar velocity flows.  

Our demonstration study of three clusters (Virgo, A1459, and A1066) showed all three have tilted, infall-caustic-like shapes in $\mbox{S}_v$. If these are infall artifacts, the amount of tilt seen in the more distant clusters lets us rule out transverse velocity between observer and cluster as the sole cause of the tilt.  However, shear flows (rotation) remain a plausible possibility as the estimated velocities needed are all reasonable.  We note that Virgo shows both a pronounced tilt and a pronounced caustic-like shape in $\mbox{S}_v$ which is not matched on one side by SIM but which could be matched, perhaps, by some model incorporating shear flow.

Since motion towards or away from a nearby cluster causes the artifact's apparent width  to shrink or grow, it may be possible to find our infall velocity towards Virgo by comparing Virgo to distant clusters.  
Application of this method to our three clusters, as a demonstration, gave promising results if the length in the ratio $\mathcal{W}$ is determined by directly measured finger length (which probably reflects dispersion of late type spirals)  rather than by virial dispersion of all galaxies in the cluster core. 

\paragraph*{Future work:}
As mentioned in \S\ref{sec:Red}, we have done studies of other models, such as toy models incorporating rotational flows and a SIM-based filament model, and we plan to present those results in a follow-up paper since there was not room for them here.  

Additional studies that would be straightforward to carry out include applying SIMs to a large sample of clusters at varying redshifts and directions; comparing different types of SIMs such as Yahil approximation and PSM;  comparing Virgo and SIM-based simulation in real and redshift space;  systematically searching cosmological n-body simulations for structures resembling tilted infall artifacts;
and improving the NDM algorithm for fitting envelopes.   
We have done work on some of these but welcome other researcher's contributions or ideas.

We also suggest more studies be done of $\lambda$CDM n-body simulations to find out if a constrained simulation of the Local Supercluster (such as in Klypin et al. 2003) can produce a better match to the observed redshift space shape and velocity field around Virgo  than the cosmological simulations we explored for this paper.  In addition, Ibata et al.'s (2013) discovery of an unexpected co-rotating disk of satellite galaxies about the Andromeda galaxy may be relevant to studies of flow on larger scale.

Packages of slicing software were coded in Mathematica and Matlab frameworks and are available for anyone. For more details please contact 
elizabeth.praton@fandm.edu and mhamdy@nriag.sci.eg.

\section* {Acknowledgements}
We thank Lindsey Mahovetz for contributing to this study by exploring A1459 and A1066 and doing preliminary envelope fits, and Bingxin Zhang for exploring clusters in the GIF, Bolshoi, and Multi-Dark 1 n-body outputs.  Also, we would like to thank Prof. Ravi Sheth for providing us with an N-body simulation and useful discussion.  Finally, we thank the reviewer for suggestions which improved this paper.

{}

\appendix

\renewcommand{\theequation}{\Alph{section}\arabic{equation}}
\renewcommand{\thefigure}{\Alph{section}\arabic{figure}}

\section{Shells \& Envelope in 2D}
\label{app:2DEnv}
\setcounter{equation}{0}
\setcounter{figure}{0}

In this paper,  
we work in the local universe, where a galaxy's observed velocity is the vector sum of the Hubble velocity and peculiar velocities of galaxy and observer. 
This approximation holds for redshifts $z \lesssim 0.1$ (e.g.~\S\ref{sec:locUniv} below).

\subsection{Local Universe Approximation}

\label{sec:locUniv}

By convention, a galaxy's observed speed is defined $s_{obs} \equiv c z$ where $c$ is speed of light and $z$ is observed redshift.  If observer and galaxy have non-relativistic peculiar velocities with line-of-sight components $v_0$  and $v_G$ respectively, and if redshift due to expansion of universe is $z_R$, then $1 + z = (1 + v_G/c)(1 + z_R)(1 - v_0/c)$ (Harrison 1974).  

To second order in $z_R$, luminosity distance is $d_L \simeq (c z_R/H_0) [1 + (1/2)(1 - q_0) z_R]$, where $q_0 = \Omega_{m,0}/2 - \Omega_{\Lambda,0}$ and $\Omega_{m,0}$ and $\Omega_{\Lambda,0}$ are current values of closure constants due to matter and cosmological constant, respectively (e.g., Carroll \& Ostlie 2007).

Combining $d_L$ and $s_{obs} = c z = c z_R + (1 + z_R)(v_G - v_0)$ yields, to first order in $z_R$,
\beq
s_{obs} \simeq H_0 d_L ( 1 - n z_R) + (v_G - v_0) (1 + z_R),
\eeq
where $n \equiv (1/2)(1 - q_0)$.  The values of $n$ for the concordance model ($\Omega_{\Lambda,0} = 0.7$, $\Omega_{m,0} = 0.3$),  empty model ($\Omega_{\Lambda,0} = 0$, $\Omega_{m,0} = 0$), and flat matter-only model ($\Omega_{\Lambda,0} = 0$, $\Omega_{m,0} = 1$) are n = 0.8, 0.5, and 0.25, respectively.

Thus, if we restrict ourselves to galaxies with observed redshift $z \lesssim 0.1$ and set $r = d_L$, observed speed can be approximated with less than 10\% error as the line of sight component of the vector resulting from the sum of the Hubble flow $\mathbf{v}_H$ and the peculiar velocity $\mathbf{v}_{pec}$ minus the observer's velocity $\mathbf{v}_0$: 
\beq
s_{obs} \simeq H_0 r + (v_G - v_0) = \mathbf{e}_r \cdot (\mathbf{v}_H + \mathbf{v}_{\text{pec}} - \mathbf{v}_0). 
\label{eq:approxsobs}
\eeq


\subsection{Shells in x-y plane} \label{app:2Dsh}

Let X be a coordinate system centered on the observer, where the distance $r$ and position angles $\theta$ and $\phi$ are related to $x$, $y$, and $z$ by $x = r\sin\theta\cos\phi$,$y = r\sin\theta\sin\phi$, $z = r\cos\theta$.

\begin{figure} 
   \centering \vspace{-0.3cm}
\includegraphics[width=4.5cm,height=3.5cm]{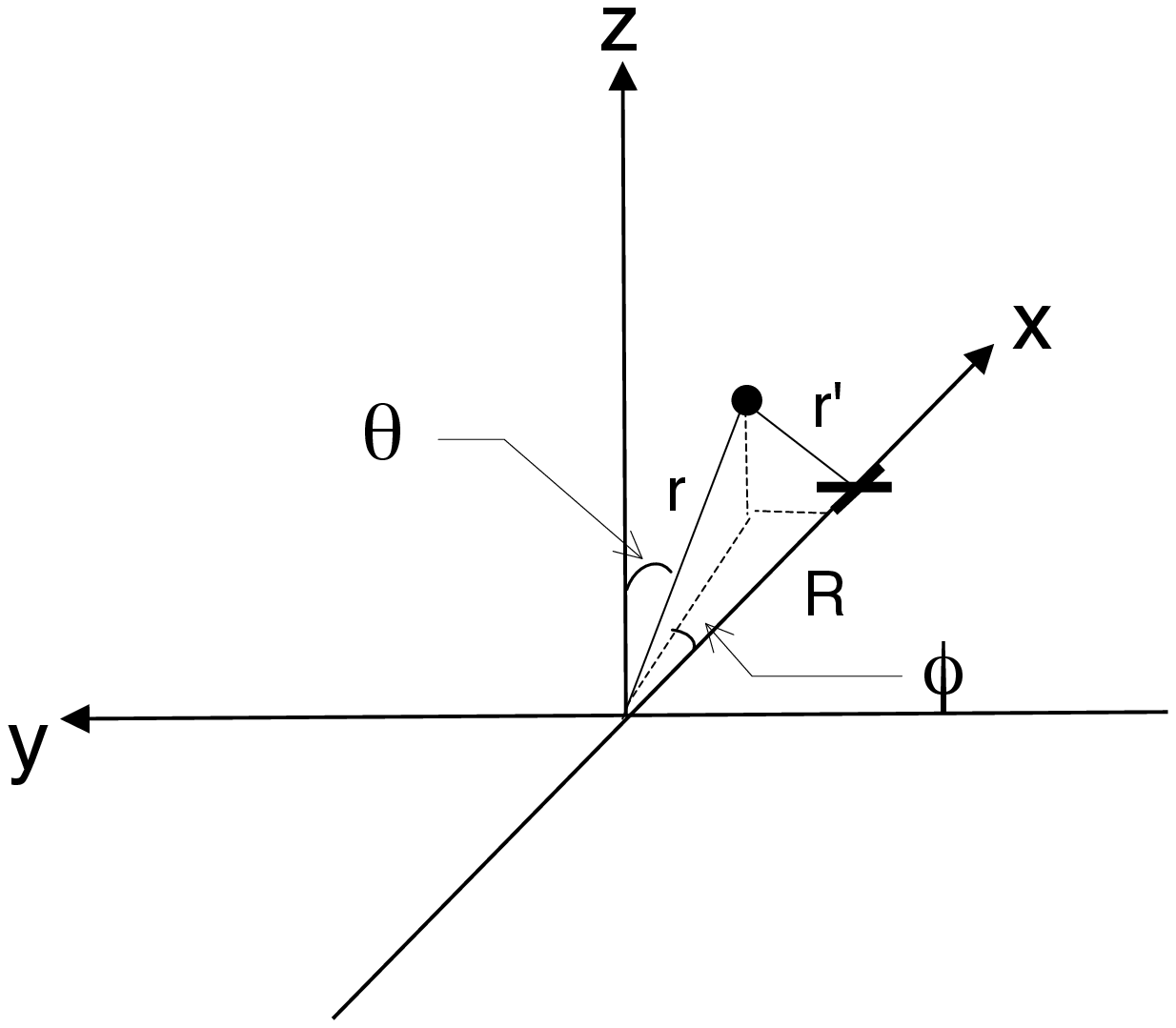} \hspace{0.5in} 
   \caption{Coordinate system useful for 2D envelope.  ({\em Left}) Galaxy ({\em point}) at arbitrary location. 
   Cluster center ({\em cross}) lies along x-axis. ({\em Right}) Galaxy in x-y plane. Angular position is 
   specified by azimuthal angle $\phi$.}
   \label{fig:thetaphi}
\end{figure}

Let the cluster be centered at ${\bf x} = R {\bf e}_x$, thus giving the cluster core longitude and latitude coordinates $l = 0$, $b = 0$. Let  $\mbox{X}'$ be a coordinate system centered on the cluster, and let the $z$ and $z'$ axes be parallel. (See Fig.~\ref{fig:thetaphi}.)

The Hubble flow is defined ${\bf v}_H \equiv H_0 r {\bf e}_r$. (Note that the reference point for the Hubble flow is immaterial, i.e., we could also write ${\bf v}_H = H_0 r' {\bf e}_{r'}$ if we wish).

The peculiar velocity field is defined as the difference between the actual velocity field and the Hubble flow. Consider a peculiar velocity field centered on the cluster core which has an inward directed radial component with magnitude $s_{\text{rad}}({\bf x}')$ and  a rotation component $s_{\text{rot}}({\bf x}')$ about the $z'$ axis. Then the peculiar velocity field may be written ${\bf v}_{\text{pec}} = -s_{\text{rad}} {\bf e}_{r'} + s_{\text{rot}} {\bf e}_{\phi'}$, where the unit vectors ${\bf e}_{r'}$ and ${\bf e}_{\phi'}$ written in terms of $x$, $y$, and $z$ are
\begin{equation}
{\bf e}_{r'} = \frac{(x - R){\bf e}_x + y{\bf e}_y + z{\bf e}_z}
	{((x - R)^2 + y^2 + z^2)^{1/2}}, \qquad
{\bf e}_{\phi'} = \frac{-y{\bf e}_x + (x - R){\bf e}_y}
	{((x - R)^2 + y^2)^{1/2}}.
\end{equation}
The observer may or may not obey the infall law; let ${\bf v}_0 = v_{0x} {\bf e}_x + v_{0y} {\bf e}_y + v_{0z} {\bf e}_z$ be the arbitrary peculiar velocity of the observer.

In the local universe (Eq.~\ref{eq:approxsobs}), the observed speed of any galaxy in the infall field is 

$s_{\text{obs}} = {\bf e}_r \cdot({\bf v}_H + {\bf v}_{\text{pec}} - {\bf v}_0) =
 H_0 r \, {\bf e}_r \cdot {\bf e}_r - 
	s_{\text{rad}}\, {\bf e}_{r'} \cdot {\bf e}_r +
	s_{\text{rot}}\, {\bf e}_{\phi'}\cdot {\bf e}_r - {\bf v}_0 \cdot {\bf e}_r$,
or
\begin{align}	 \label{eq:speedobs}
    s_{\text{obs}} &= H_0 r - s_{\text{rad}} \frac{(r - R \sin\theta\cos\phi)}{r'}
		- s_{\text{rot}} \frac{R \sin\theta\sin\phi}{(r'^2 - r^2\cos^2\theta)^{1/2}} \nonumber \\
   &  \qquad 	- (v_{0x}\sin\theta\cos\phi + v_{0y}\sin\theta\sin \phi	+ v_{0z} \cos \theta). 
\end{align}
where the distance to the cluster center is $r' = (r^2 + R^2 - 2 r R \sin\theta\cos\phi)^{1/2}$.

We may 
simplify
Eq.~\ref{eq:speedobs} by considering just the $x$--$y$ plane, 
for which $\theta = 90^\circ$. Then

\beq
s_{\text{obs}} = H_0 r - s_{\text{rad}}\frac{r - R\cos\phi}{r'} - s_{\text{rot}} \frac{R\sin\phi}{r'}
	- v_{0x} \cos\phi - v_{0y} \sin\phi.
	\label{eq:sobs1}
\eeq

Consider the cross-sections of a nested set of shells of radius $r'$ in the $x$--$y$ plane. Assume that the infall velocity field depends only on radius $r'$ and the rotation field (if any) is symmetric about the $z'$ axis. Then, by substituting 
$r = R\cos\phi \pm r'(1 - (R/r')^2\sin^2\phi)^{1/2}$ into Eq.~\ref{eq:sobs1}, we see that any given shell of radius $r' = constant$ has an observed velocity 
\begin{align}		\label{eq:sobsxy} 
s_{\text{obs}} &= (H_0 R - v_{0x})\cos\phi - \left(s_{\text{rot}}(r') \frac{R}{r'} + v_{0y}\right)\sin \phi  \nonumber \\
	&  \qquad  \pm \left(H_0 r' - s_{\text{rad}}(r')\right) \left(1 - \left(\frac{R}{r'}\sin\phi\right)^2\right)^{1/2},
\end{align}
where the positive sign is for the far side of the shell and the negative sign in for the near side of the shell. 

\subsection{Envelope in x-y plane} \label{app:2den}

A continuous distribution of shells from the virial radius to the turnaround shell distort in redshift space into 
a rhomboid shaped structure.  We can find the envelope bounding the structure (as done by RG89) 
by considering the shells inside the turnaround shell as a family of curves parametrized by the shell radius $r'$.  

According to Courant (1936), we obtain the envelope of a family of curves $f(x,y,c) = 0$ by considering the 
two equations $f(x,y,c) = 0$  and $\pa f/\pa c = 0$ simultaneously and attempting to either eliminate $c$ or 
express $x$ and $y$ as functions of $c$.  

In our case, $r'$ is the constant $c$, so we take a partial derivative of Eq.~\ref{eq:sobsxy} with respect to $r'$. 
The result is
\begin{align}  \label{eq:pasobsxy}
\frac{\pa s_{\text{obs}}}{\pa r'} &= 
	\left(\frac{s_{\text{rot}}}{r'~} - \frac{ds_{\text{rot}}}{dr'~}\right) \frac{R}{r'}
	\sin\phi \pm \frac{1}{(1 -  (R/r')^2 \sin^2\phi)^{1/2}} \times 	\nonumber
\\
& \qquad   \left[H_0 - \frac{ds_{\text{rad}}}{dr'~} + 
	\left(\frac{ds_{\text{rad}}}{dr'~} - \frac{s_{\text{rad}}}{r'}\right)
	\left(\frac{R}{r'}\right)^2 \sin^2\phi \right].
\end{align}

Note we allow for the possibility of rotation, although we assume that in the $x$--$y$ plane the rotational 
speed $s_{\text{rot}}$, like the radial speed $s_{\text{rad}}$, depends only on the shell radius $r'$.

To solve for the equation of the envelope, we set $\pa s_{\text{obs}}/\pa r' = 0$ and solve Eq.~(\ref{eq:pasobsxy}) 
for $(R/r')\sin\phi$; then substitute that expression into Eq.~(\ref{eq:sobsxy}). The net result is a parametric 
equation for the envelope:  $s_{\text{obs}} = s_{\text{obs}}(r')$ and $\phi = \phi(r')$.  

Explicitly, if we introduce a shorthand notation $u \equiv (R/r')\sin\phi$ and
\begin{equation}\begin{split}
A(r') \equiv \frac{s_{\text{rot}}}{r'} - \frac{ds_{\text{rot}}}{dr'~}, \\ \qquad
B(r') \equiv \frac{s_{\text{rad}}}{r'} - \frac{ds_{\text{rad}}}{dr'~}, \\ \qquad 
C(r') \equiv H_0 - \frac{ds_{\text{rad}}}{dr'~},
\label{eq:ABC}
\end{split}\end{equation}
then $\pa s_{\text{obs}}/\pa r' = 0$ becomes
\begin{equation}
A u (1 - u^2)^{1/2} = \mp(C - B u^2)
\label{eq:ABCu}
\end{equation}
where minus is for the far side of the shells and plus is for the near side.
Solving for $u$
yields four solutions:  $u = u_+$, $u = -u_+$, $u = u_-$, $u = -u_-$, where
\beq
u_+ = (a + b)^{1/2}, \qquad   u_- = (a - b)^{1/2}
\label{eq:u+u-}
\eeq
and where $a$ and $b$ are defined in terms of the parameters $A$, $B$, $C$ as 
\beq
a \equiv \frac{2 B C + A^2}{2 (A^2 + B^2)}, \quad 
	b \equiv \frac{|A|(A^2 + 4(B - C)C)^{1/2}}{2(A^2 + B^2)}.
\eeq

On any given edge of the envelope, two of the four solutions are spurious, introduced by the squaring of  Eq.~(\ref{eq:pasobsxy}).  To determine which are valid, we proceed as follows.

Look at the special case $A = B = 1$ (for convenience) and $C = 1 - \delta$, where $\delta \ll 1$.  This is the region just inside the turnaround radius.  We can't use the turnaround region itself ($B = C$) because the solutions have a singularity there.  If we expand in powers of $\delta$, the valid solutions are the ones which yield $\pa s_{\text{obs}}/\pa r' = 0$, i.e., that satisfy Eq.~(\ref{eq:ABCu}).  

The result reveals that  the 2D x-y plane envelope bounding the structure produced in redshift space by the family of shells lying within the turnaround radius is given by the following set of parametric equations.

For any given shell of radius $r'$, the azimuthal angle $\phi$ of the redshift-space envelope at the spots tangent to the shell's near and far sides are
\begin{align}
\mbox{(near)} \qquad
\sin\phi &= \left\{ \begin{array}{ll}
			\frac{r'}{R} u_- & \mbox{ if $\phi > 0$,}\\
			-\frac{r'}{R} u_+ & \mbox{ if $\phi < 0$,}
		   \end{array}
	   \right. 
\label{eq:envelopesinphi}
\\
\mbox{(far)} \qquad
\sin\phi &= \left\{ \begin{array}{ll}
			\frac{r'}{R} u_+ & \mbox{ if $\phi > 0$,}\\
			-\frac{r'}{R} u_- & \mbox{ if $\phi < 0$,}
		   \end{array}
	   \right. \nonumber
\end{align}
where $u_+$ and $u_-$ are given by Eq.~(\ref{eq:u+u-}). 

The associated observed speed $s_{\text{obs}}$ is found by substituting the value of $\sin\phi$ into Eq.~(\ref{eq:sobsxy}).  Note that $r'$ is a {\em parameter}. Let it vary from the virial radius to the turnaround radius to find the corresponding $\phi$ and $s_{\text{obs}}$ of the spots on the bounding envelope tangent to each of the shells in redshift space and then use $x = s_{\text{obs}} \cos\phi$, $y = s_{\text{obs}} \sin\phi$ to plot the envelope.  

Also note that each shell turns inside out in redshift space, so the envelope edge tangent to galaxies that are physically nearer to the observer (shells' {\em near} sides) actually lies on the {\em far} side in redshift space, and vice versa.

\section{Shells and Envelope in 3D}
\label{app:3DEnv}
\setcounter{equation}{0}
\setcounter{figure}{0}

\subsection{Shells in 3D}\label{app:3dsh}

The derivation for a 3D infall envelope in redshift space is similar to that for the 2D envelope, but we will make some changes to the coordinate system.  Also, to simplify the derivation, we will not include rotational flow.

Let X and $\mbox{X}'$ be coordinate systems centered on the observer and on the cluster, respectively, and let the cluster lie on the x axis at ${\bf x} = R {\bf e}_z$, as in the 2D derivation 
(Appendix~\ref{app:2DEnv}).  

\begin{figure}\centering \vspace{-0.2cm}
   \includegraphics[width=5cm,height=3.5cm]{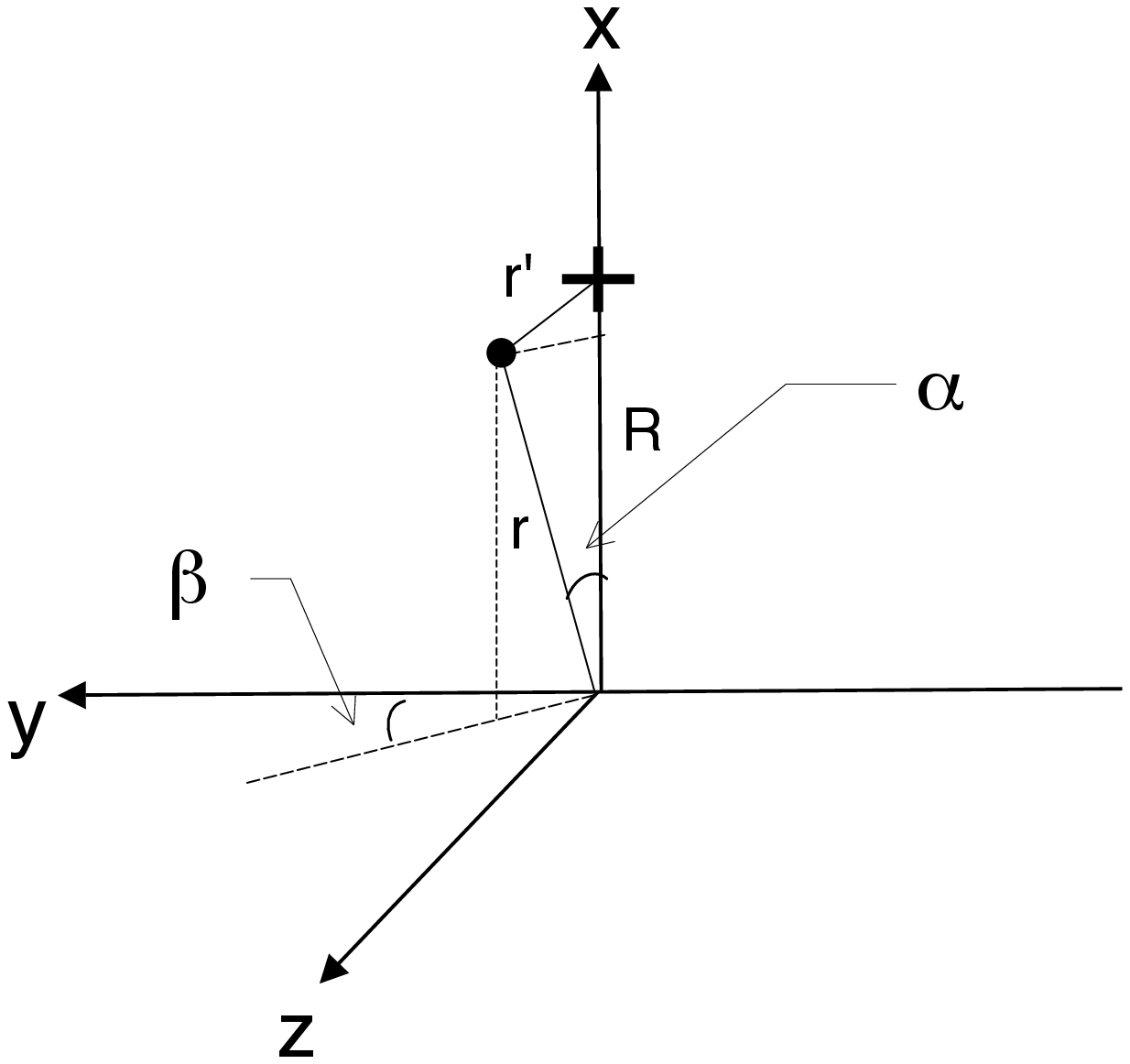} \vspace{-0.3cm}
   \caption{ Coordinate system useful for 3D envelope.   Cluster center ({\em cross}) still lies on x-axis, but angular position of a galaxy ({\em point}) is specified by  polar angle $\alpha$ and azimuthal angle $\beta$ defined with respect to cluster axis, not z-axis. (Compare with Fig.~\ref{fig:thetaphi}.) }  
   \label{fig:alphabeta}
\end{figure}

However, for convenience, make a change in angular coordinates.  In place of  
$\theta$ and $\phi$, define new 
coordinates $\alpha$ and $\beta$, where $\alpha$ is a polar angle measured from the x-axis and $\beta$ is an azimuthal angle measured about the x-axis in the y-z plane, so that $x = r \cos\alpha$, $y = r \sin\alpha \cos\beta$, and $z = r \sin\alpha \sin\beta$, where $r$ is the distance from the observer 
(see Fig.~\ref{fig:alphabeta}).
These coordinates are useful because letting $\beta$ vary from 0 to $360^\circ$ will generate the 3D surface of the envelope.   

Consider a point on a shell a distance $r'$ from the cluster center.  Assume there is no rotational flow, and peculiar velocity near the cluster is spherically symmetric and radial with respect to the cluster center.  Since there is no rotational flow, for convenience orient the y-axis in the direction of the observer's transverse motion with respect to the cluster, so the observer's peculiar velocity ${\bf v}_0$ has no component in the z-direction:    ${\bf v}  _0 = v_{0x} {\bf e}_x + v_{0y} {\bf e}_y $.

As before, the point's velocity with respect to the cluster center is ${\bf v}   = (H_0 r' - s_{\text{rad}}(r')) {\bf e}_{r'}$, where $H_0$ is the Hubble constant and $s_{\text{rad}}$ is inward directed radial peculiar velocity.  Thus, its velocity with respect to the observer is ${\bf v}  _{\text{rel}} = H_0 R \, {\bf e}_x + (H_0 r' - s_{\text{rad}}(r')) \, {\bf e}_{r'} - {\bf v}_0$
and the observed speed of the point is $s_{\text{obs}} = {\bf v}  _{\text{rel}} \cdot {\bf e}_r$ or
\begin{equation}
s_{\text{obs}} = (H_0 R) \, {\bf e}_x \cdot{\bf e}_r + (H_0 r' - s_{\text{rad}}(r')) \, {\bf e}_{r'} \cdot {\bf e}_{r} - {\bf v}  _0 \cdot {\bf e}_r .
\label{eq:sobs}
\end{equation}

Define $\gamma$ as the angle between ${\bf e}_{r'}$ and ${\bf e}_r $ so ${\bf e}_{r}\cdot {\bf e}_{r'} = \cos\gamma$ .  Then by law of sines, $(\sin\gamma)/R = (\sin\alpha)/r'$ (see Fig.~\ref{fig:alphabeta}) and 
\begin{equation}
{\bf e}_{r}\cdot {\bf e}_{r'} = \pm\sqrt{1 - \left(\frac{R}{r'}\right)^2 \sin^2 \alpha},
\label{eq:rdotr}
\end{equation}
where (+)  is for points on the far side of the shell and (-) is for near side, since $\cos\gamma$ is positive on far side and negative on the near side.

Plugging Eq.~(\ref{eq:rdotr}) and  ${\bf e}_{r}= \cos\alpha \, {\bf e}_x + \sin\alpha \cos\beta \, {\bf e}_y  + \sin\alpha \sin\beta \, {\bf e}_z$ into Eq.~(\ref{eq:sobs}) yields the observed speed of a point that lies at polar angle $\alpha$ and azimuthal angle $\beta$ on a shell of radius $r'$:
\begin{equation}\begin{split} 
s_{\text{obs}} = (H_0 R - v_{0x}) \cos\alpha - v_{0y} \sin\alpha \cos\beta \pm \\ \qquad
(H_0 r' - s_{\text{rad}}(r')) \sqrt{1-\left(\frac{R}{r'}\right)^2 \sin^2\alpha},
\label{eq:sobs2}
\end{split}\end{equation}
where (+) is for the far side and (-) is for the near side.

\subsection{3D Envelope}\label{app:3den}

To obtain the envelope of the family of shells lying within turnaround, take the partial derivative with respect to $r'$.  Then set $\partial s_{\text{obs}}/\partial r' = 0$ and express $s_{\text{obs}}$ and $\phi$ as functions of $r'$.  After some massage, we get
\begin{equation}
\frac{\partial s_{\text{obs}}}{\partial r'} = \pm \frac{1}{(1 - u^2)^{1/2} } (C - B u^2) = 0,
\end{equation}
where (+) is for the near side and (-) is for the far side, and $u$, $B$, and $C$ are defined as before:  $u \equiv (R/r') \sin\alpha$, $B \equiv s_{\text{rad}}/r' - d s_{\text{rad}}/dr'$, and $C \equiv H_0 - d s_{\text{rad}}/dr'$.

As long as $u \neq \pm 1$, then $C-Bu^2 = 0$ so $u =\pm\sqrt{C/B}$.  
Note that at turnaround, $B = C$, since $s_{\text{rad}} = H_0 r'$ there.  So $u = \pm 1$ at turnaround where solution becomes invalid.  Inside turnaround, $B > C$ (since $s_{\text{rad}} > H_0 r'$) and $u < \pm 1$.

Plugging in definition of $u$ and noting that $\sin\alpha$ is positive for the entire range of $\alpha$ (0 to $180^\circ$) gives the result $\sin\alpha = + (r'/R)\sqrt{C/B}$.  

Here, $r'$ is the parameter.  Let it vary to find out what $\alpha$ is at the tangent points of each shell inside the turnaround region.  Plug this back into Eq.~(\ref{eq:sobs2}) and let $\beta$ range from 0 to $360^\circ$ to generate the surface.

The set of parametric equations for generating the polar angle $\alpha$ and the observed speed $s_{\text{obs}}$ of points on the 3D envelope for azimuthal angle $\beta$ ranging from 0 to $360^\circ$ is thus
\begin{equation}
\sin\alpha = \frac{r'}{R} \sqrt{\frac{C}{B}},
\label{eq:3DenvAlpha}
\end{equation}

\begin{equation}  \label{eq:3DenvSobs} \begin{split} 
s_{\text{obs}} &= (H_0 R - v_{0x}) \sqrt{1-\left(\frac{r'}{R}\right)^2 \frac{C}{B} } - v_{0y} \frac{r'}{R} \sqrt{\frac{C}{B}} \cos\beta \pm \\ & \quad
(H_0 r' - s_{\text{rad}}(r')) \sqrt{1-\frac{C}{B}}, 
 \end{split} \end{equation}
where
\begin{equation}
B \equiv \frac{s_{\text{rad}}}{r'} - \frac{d s_{\text{rad}}}{dr'}, \qquad C \equiv H_0 - \frac{d s_{\text{rad}}}{dr'},
\end{equation}
 and $r'$ is a parameter ranging from the virial radius $r_{\text{vir}}$ to the turnaround radius $r_{\text{turn}}$ and (+) is for points on the shells' far sides and (-) for points on the shells' near sides.  As before,  points physically closer to the observer (shells' {\em near} sides) are on envelope's {\em far} side in redshift space, and vice versa.

This equation gives the 3D envelope when there is no rotational flow $s_{\text{rot}} = 0$ and axes are set up so observer's motion relative to the cluster is ${\bf v}  _0 = v_{0x} {\bf e}_x + v_{0y} {\bf e}_y $. 

\section{Width-to-Length Ratio in PSM}
\label{app:TurnWid}
\setcounter{equation}{0}
\setcounter{figure}{0}
\setcounter{table}{0}

The Praton-Schneider Model (PSM) for cluster infall is based on spherical accretion onto a mass seed in an otherwise uniform and expanding matter dominated universe with no cosmological constant.  The equations of motion are well known (see, e.g., Peebles 1980), but the model is modified to include a `virialized' region and is parametrized in terms of the cluster's virial dispersion and angular size of its turnaround region, along with $\Omega_0$, as detailed in the appendix of PS94.

So, what intrinsic ratio $\mathcal{W}$ do we expect in PSM for various values of $\Omega_0$ in a simple matter-dominated universe with no cosmological constant?

In the model, the virial speed $s_{\text{vir}}$ is related to the observed dispersion $\sigma_{\text{vir}}$ in the usual way:
$
s_{\text{vir}} = \sqrt{3} \sigma_{\text{vir}}.
$
The radius $r_{\text{vir}}$ of the virialized region in the model is defined such that $s_{\text{vir}} = \sqrt{GM_{\text{vir}}/r_{\text{vir}}}$ where $M_{\text{vir}}$ is the mass inside the radius. This definition is a simple approximation in which the kinetic and potential energies of the shell bounding the virialized region are assumed to obey the virial theorem (Primack 1984, PS94).  

Combining the relations between $s_{\text{vir}}$ and $r_{\text{vir}}$ (Eq.~A11 in PS94) and $r_{\text{vir}}$ and $r_{\text{turn}}$ (Eq.~A12 in PS94) that result from the above definition yields the following expression for the ratio $\mathcal{W}_0$ of the intrinsic width $H_0 \, r_{\text{turn}}$ of the redshift space artifact  to its length $s_{\text{vir}}$:
\begin{equation}
\mathcal{W}_0 \equiv
\frac{H_0\, r_{\text{turn}}}{s_{\text{vir}}} = \frac{1}{\pi^{2/3}(3\pi/2 + 1)^{1/3}} \frac{\Omega_0\, f(\Omega_0)}{(1 -\Omega_0)^{3/2}} g(\Omega_0).
\label{app_eq:widtolen}
\end{equation}
The function $f(\Omega_0)$ (Eq. A6 in PS94) is
\begin{equation}
f(\Omega_0) \equiv 2\frac{\sqrt{1 - \Omega_0}}{\Omega_0} - \cosh^{-1}\left(\frac{2 - \Omega_0}{\Omega_0}\right).
\end{equation}
and the function $g(\Omega_0)$ is 
\begin{equation}
g(\Omega) \equiv \left[\frac{M(\pi)}{M_{\text{vir}}}\right]^{1/3} 
= \left[\frac{f(\Omega_0)^{2/3} + (3\pi/2 + 1)^{2/3}}{f(\Omega_0)^{2/3} + \pi^{2/3}}\right]^{1/3},
\end{equation}
where $M(\pi)$ is the mass inside the turnaround radius.

Equation~\ref{app_eq:widtolen} has the following values in the limits $\Omega_0 \rightarrow 0$ and $\Omega_0 \rightarrow 1$:
\begin{equation}
\lim_{\Omega_0 \to 0}  \mathcal{W}_0
= \frac{2}{\pi^{2/3} (3\pi/2+1)^{1/3}} \approx 0.522
\end{equation}
and
\begin{equation}
\lim_{\Omega_0 \to 1} \mathcal{W}_0 
= \frac{4/3}{\pi^{8/9} (3\pi/2+1)^{1/9}} \approx 0.397.
\end{equation}
Table~\ref{tab:intrinsicW} gives values of the ratio for other values of $\Omega_0$.
%


\begin{table}
\caption{\small Intrinsic width to length ratios $\mathcal{W}_0$ of redshift-space infall artifact in Praton-Schneider model.}
{\small
\begin{center}
\begin{tabular}{l|cccccc}
\hline
\hline
$\Omega_0$ & 0 & 0.1 & 0.2 & 0.3 & 0.4 & 0.5  \\
$\mathcal{W}_0$  & 0.522 & 0.487 & 0.468 & 0.454 & 0.442 & 0.432  \\
\hline
$\Omega_0$  & 0.6 & 0.7 & 0.8 & 0.9 & 1 & \\
$\mathcal{W}_0$   & 0.424 & 0.416 & 0.409 & 0.403 & 0.397 & \\

\hline
\hline
\end{tabular}
\end{center}
}
\label{tab:intrinsicW}
\end{table}%

\end{document}